\documentclass[a4paper,11pt,nohyper]{article}
\pdfoutput=1
\usepackage{jheppub}

\usepackage[T1]{fontenc}
\usepackage[mathletters]{ucs} 
\usepackage[utf8x]{inputenc}
\usepackage{multirow,booktabs,array}
\usepackage{amssymb}
\usepackage{amsmath}
\usepackage{lmodern}
\usepackage{tikz}
\usepackage{subfig}
\usepackage{xcolor}
\usepackage{braket}
\usepackage{verbatim}
\usepackage{multirow}
\usepackage{graphics,graphicx}
\usepackage{psfrag}
\usepackage{empheq} 
\usepackage{mathtools}
\usepackage{cleveref}

\def\n{\nu}
  
  \let\n=\nu

 \def\bd{\begin{document}} \def\ed{\end{document}}
\def\ds{\documentstyle} \let\fr=\frac \let\bl=\bigl \let\br=\bigr
\let\Br=\Bigr \let\Bl=\Bigl
\let\bm=\bibitem
\let\na=\nabla

\newcommand{\boxedeq}[2]{\begin{empheq}[box={\fboxsep=6pt\fbox}]{align}\label{#1}#2\end{empheq}}
\newcommand{\be}{\begin{equation}}
\newcommand{\ee}{\end{equation}}

\newcommand{\bea}{\begin{eqnarray}}
\newcommand{\eea}{\end{eqnarray}}

\definecolor{ao(english)}{rgb}{0.0, 0.5, 0.0}

\usepackage{amsmath,latexsym,amssymb,slashed}
\usepackage{mathrsfs}

\title{Renormalized entanglement entropy and curvature invariants}
 \author{Marika~Taylor}
  \author{and Linus~Too}
 \affiliation{School of Mathematical Sciences and STAG Research Centre, University of Southampton\\Highfield, Southampton, SO17 1BJ, UK.} 
\bigskip
\emailAdd{m.m.taylor@soton.ac.uk, l.h.y.too@soton.ac.uk}

\abstract{Renormalized entanglement entropy can be defined using the replica trick for any choice of renormalization scheme; renormalized entanglement entropy in holographic settings is expressed in terms of renormalized areas of extremal surfaces. In this paper we show how holographic renormalized entanglement entropy can be expressed in terms of the Euler invariant of the surface and renormalized curvature invariants.  For a spherical entangling region in an odd-dimensional CFT, the renormalized entanglement entropy is proportional to the Euler invariant of the holographic entangling surface, with the coefficient of proportionality capturing the (renormalized) F quantity. Variations of the entanglement entropy can be expressed elegantly in terms of renormalized curvature invariants, facilitating general proofs of the first law of entanglement.
}

\preprint{}

\begin{document}
	\flushbottom
	\maketitle
	
	
	
	
	\section{Introduction and summary}

Viewed from the perspective of quantum field theory, entanglement entropy is an unusual quantity. Entanglement entropy is usually expressed as a regulated quantity, with the regulator being a short distance cutoff but the regulated power law divergences depend on the details of the regulation scheme. Accordingly the main focus is on the so-called universal terms, the coefficients of logarithmic divergences, as these are related to the coefficients of the Weyl anomaly of the stress energy tensor. 

For condensed matter and quantum information applications, quantum field theory is used as an intermediate tool to describe a system with an inherent lattice cutoff. In such contexts the short distance regulator has a physical interpretation as the lattice spacing. 
If quantum field theory is used to describe a continuum system, there is no inherent physical cutoff: in quantum field theory we work with renormalized quantities, rather than regulated quantities. Renormalized entanglement entropy has been developed in \cite{Taylor:2016aoi,Taylor:2017zzo,Anastasiou:2017xjr,Anastasiou:2018rla,Anastasiou:2018mfk,Anastasiou:2019ldc,Anastasiou:2020smm}. 

The focus in this paper will be on the holographic definition of renormalized entanglement entropy in terms of the renormalized area of entangling surfaces, as shown in \eqref{area2} and \eqref{renorma}.
 Renormalized entanglement entropy can however be defined in generality using the replica approach, which is in practice almost always used for explicit computations of entanglement entropy in quantum field theory, see for example \cite{Calabrese:2007mtj,Calabrese:2009qy,Casini:2009sr}. The bare entanglement entropy is expressed as
\be
S = - {\rm Lim}_{n \rightarrow 1} \left ( \partial_{n} \left [ {\rm Tr} (\rho^n) \right ] \right )
\ee
where $\rho$ is the density matrix of the (reduced) state. This expression can be written in terms of partition functions as
\be
S = - {\rm Lim}_{n \rightarrow 1} \left ( \partial_{n} \left [ Z(n) - n Z(1) \right ] \right )
\ee
where $Z(1)$ denotes the partition function and $Z(n)$ denotes the partition function on the replica space ($n$ copies of the original space joined together cyclically). The renormalized entanglement entropy can then be defined as 
\be
S_{\rm ren} = - {\rm Lim}_{n \rightarrow 1} \left ( \partial_{n} \left [ Z_{\rm ren}(n) - n Z_{\rm ren}(1) \right ] \right ).
\ee
Here the partition function $Z_{\rm ren}(1)$ is renormalized using any method of renormalization. The partition function on the replica space 
inherits the same UV divergence structure and thus the renormalized $Z_{\rm ren}(n)$ can be defined without ambiguities from the original renormalization scheme. 

In \cite{Page:1993wv} Page characterised information recovery from black holes in terms of the time dependence of the entanglement entropy of the Hawking radiation. A number of recent works, such as \cite{Almheiri:2019hni,Almheiri:2019yqk,Almheiri:2019qdq}, have discussed how the Page curve for Hawking radiation can be recovered from semiclassical geometry. It is interesting to note that these discussions inherently rely on a finite (renormalized) notion of entanglement entropy, as defined above.

\bigskip

The UV divergences in the bare entanglement entropy are associated physically with local entanglement at the boundary of the entangling region. The renormalized entanglement entropy is instead associated with non-local entanglement between the entangling region and its complement. The behaviour of renormalized entanglement entropy in various phases of holographically realised quantum field theories was explored in  \cite{Taylor:2017zzo}. 

Renormalized entanglement entropy is computed holographically in terms of the renormalized area of minimal surfaces. The latter topics has been explored right from the very early days of the AdS/CFT correspondence \cite{Henningson:1999xi, Graham:1999pm}, as it is also relevant to the holographic computation of Wilson loops. Within the mathematics community, there has been considerable study of renormalized areas of surfaces, see for example \cite{Alexakis:2010zz,Gover:2016xwy,Gover:2016hqd,Zhang:2017lcd,Graham:2016jqn,Graham:2019epd}. Connections between renormalized areas, entanglement and the Willmore functional have been explored within both the mathematics and the physics communities 
\cite{Graham:2017bew,Seminara:2018pmr}. 

\bigskip

The main goal of this paper is to demonstrate how the renormalized entanglement entropy can be expressed in terms of the Euler characteristic and other conformal invariants in odd-dimensional UV conformal field theories dual to gravity in even dimensions. The restriction to even dimensions is for the usual reason: conformal field theories in even dimensions have conformal anomalies, and accordingly the renormalized entanglement entropy is not a conformal invariant. For AdS$_4$/CFT$_3$, the required geometric analysis is already contained in \cite{Alexakis:2010zz}; here we interpret these mathematics results physically, particularly in terms of the F quantity. We then generalize the approach of \cite{Alexakis:2010zz} to AdS$_6$/CFT$_5$ dualities. 

We show that the renormalized entropy $S(\Sigma)$ for a static entangling surface $\Sigma$ in an asymptotically AdS$_{2n}$ spacetime has the following structure:
\begin{equation}
S (\Sigma) \sim(-1)^{n+1}  {\cal F}_{n} \; \chi (\Sigma) - \sum_{r} {\cal W}_{r} (\Sigma) - \sum_{p} {\cal H}_{p} (\Sigma){- \sum_{q} {\cal I}_{q} (\tilde{B}).} \label{structure}
\end{equation}
In this and all subsequent expressions $S$ refers to the renormalized entanglement entropy i.e. for notational brevity we drop the subscript. The Euler invariant of the entangling surface is denoted $\chi (\Sigma)$ and ${\cal F}_{n}$ is a numerical coefficient. In everything that follows we implicitly work with spacetimes with constant negative Ricci curvature, i.e. no matter or gauge fields, but the generalization of our results to include bulk stress energy tensors would be straightforward. 

The contributions ${\cal W}_r$ are expressed in terms of the pullback of the Weyl curvature to the surface. Each such contribution is individually finite and conformally invariant; finiteness generically requires that appropriate boundary terms are included. For $n=2$ there is one single such contribution, linear in the Weyl tensor while for $n = 3$, there are two terms, linear and quadratic in the Weyl tensor. For general $n$ terms up to and including order $(n-1)$ arise. 

The contributions ${\cal H}_{p}$ are expressed in terms of scalar invariants built from the extrinsic curvature. Again, each such contribution is individually finite and conformally invariant, with boundary terms generically being required. For AdS$_{2n}$ there are contributions up to and including order $2(n-1)$ in the extrinsic curvature; all such contributions involve an even number of extrinsic curvatures.  {For $d>5$ there are ${\cal I}_{q}$ renormalized integrals containing products of Weyl and extrinsic curvature.}

The general structure of the renormalized entropy/area and the decomposition using Gauss-Codazzi relation will apply in all even dimensions. The explicit terms that arise would need to be calculated for dimensions greater than or equal to eight, and the associated positivity properties proven.

While the gravity calculation can be carried out in all even dimensions, we should note however that the quantum field theory interpretation of the results in $AdS_{2n}$ with $n \geq 4$ is unclear and there are no conformal field theories in dimension $n\geq 7$.

We note that relations between a renormalized entanglement entropy, the Euler invariant and curvature invariants has been considered in earlier works \cite{Anastasiou:2017xjr,Anastasiou:2018rla,Anastasiou:2018mfk,Anastasiou:2019ldc,Anastasiou:2020smm}. However, the underlying approach of these works is somewhat different: the renormalized entanglement entropy is not defined by using the boundary terms induced by the variational problem at the conformal boundary \cite{Papadimitriou:2005ii} as in \cite{Taylor:2016aoi,Taylor:2017zzo}, following the standard approach to holographic renormalization \cite{deHaro2001,Papadimitriou:2004ap}, but instead by adding Chern forms as boundary terms. However, the results coincide for $AAdS_4$; the Chern form and counterterm for the codimension two minimal surface renormalized area are identical as illustrated in \cite{Anastasiou:2017xjr,Alexakis:2010zz}. When the bulk entangling surface is a codimension two asymptotically hyperbolic slice of $AAdS_6$, $\Sigma=A\mathbb{H}^4$, by discarding all quantity extrinsic to $\Sigma$, the renormalized area formula $(\ref{eq:A6})$ reproduces Anderson's four dimensional renormalized volume formula \cite{Anderson:2001vr}
\begin{align}
\chi(\Sigma)=\frac{3}{4\pi^2}\mathcal{A}(\Sigma)+\frac{1}{32\pi^2}\int_{\Sigma}|W^{(4)}|^2
\end{align}
where $\mathcal{A}(\Sigma)$ is the renormalized area and $W^{(4)}$ is the Weyl tensor intrinsic to the four dimensional hyperbolic space $\Sigma$. In this case our results should coincide with \cite{Anastasiou:2018rla,Anastasiou:2018mfk,Anastasiou:2019ldc,Anastasiou:2020smm}. 

More generally, as pointed out by \cite{anastasiou2020counterterms}, the Kounterterm approach differs from the holographic renormalization procedure when the boundary Weyl tensor of the asymptotically locally $AdS$ spacetime is non-vanishing. Using the Gauss-Codazzi relations, the boundary Weyl tensor is related to projections of the bulk Weyl tensor. In $(\ref{eq:A6})$, the projections of the bulk Weyl  contributes  to the renormalized area. Hence, we anticipate that our results could  differ from the Kounterterm approach and it would be interesting to compare the results in higher dimensions.

\bigskip

The expression \eqref{structure} has several immediate physical applications. Firstly, for entangling surfaces in AdS$_{2n}$ all ${\cal W}_r$ contributions are zero, due to the vanishing of the Weyl tensor. Umbilic minimal surfaces have zero extrinsic curvature, and thus the renormalized entanglement entropy reduces to the Euler invariant term. Entangling surfaces associated with spherical entangling regions (discussed extensively in \cite{Casini:2011kv}) are indeed umbilic and thus their renormalized entanglement entropies are proportional to their Euler invariants (which are one for all $n$).
 
In \cite{Casini:2011kv} it was shown that the finite contributions to the entanglement entropy of spherical regions compute the F quantities \cite{Jafferis:2011zi} in odd dimensional conformal field theories. Renormalized entanglement entropy enables these finite contributions to be extracted elegantly, in a manifestly scheme independent manner \cite{Taylor:2016aoi, Taylor:2016kic}. By expressing the renormalized entanglement entropy in the form \eqref{structure}, it is manifest that the coefficients of proportionality ${\cal F}_{n}$ of the Euler invariants directly compute the F quantities. 

\bigskip

The second immediate application of \eqref{structure} is to variations of the entanglement entropy under changes in the background geometry (state of quantum field theory) and changes in the shape of the entangling region. The expression \eqref{structure} can be used to give an elegant proof of the first law of entanglement entropy, generalizing the work of \cite{Faulkner:2013ica} as one no longer needs to restrict to normalizable metric perturbations. 

The first variation of the entanglement entropy around spherical entangling regions in AdS takes a particularly simple and elegant form. Since such variations do not change the topology of the entangling surface, the Euler invariant contribution does not change. All contributions from the extrinsic curvature are quadratic or higher order; since the extrinsic curvature vanishes to leading order, this means the contributions ${\cal H}_p$ do not contribute to first variations (but do contribute to the second variations). By analogous reasoning, the only contribution from the Weyl terms ${\cal W}_r$ comes from the term that is linear in the Weyl tensor. Thus we arrive at 
\be
\delta S \propto \frac{-1}{4 G_{2n}} \delta {\cal W}
\ee
where $G_{2n}$ is the Newton constant and
\be
\delta {\cal W} = \int_{\Sigma} d^{2 (n-1)} x \sqrt{g} \; \delta \widetilde{W}_{1212} - \int_{\partial \Sigma} d^{2n -3} x \sqrt{h} \delta W_{1212} + \cdots
\ee
where $\delta \widetilde{W}_{1212}$ is the pullback of the normal components of the bulk linearized Weyl curvature in an orthonormal frame and
$\delta W_{1212}$  is the pullback of the normal components of the boundary linearized Weyl curvature in an orthonormal frame. The boundary terms are such that $\delta {\cal W}$ is a finite conformal invariant for a generic non-normalizable metric perturbation. Note that the boundary term vanishes 
for AdS$_4$. The ellipses denote additional boundary terms expressed in terms of higher powers of the boundary Weyl curvature that are required for $n > 3$. 

In a future work \cite{FirstLaw} we will show in detail how $\delta {\cal W}$ can be related to the renormalized stress tensor defined in \cite{deHaro2001} and hence to the variation in the energy; this gives a generalized proof of the first law \cite{Faulkner:2013ica} in a simple and elegant way. 

\bigskip

The plan of this paper is as follows. In section \ref{sec:two} we consider static entangling surfaces in asymptotically locally AdS$_4$ spacetimes; the relevant mathematical results were derived in \cite{Alexakis:2010zz}. In section \ref{sec:three} we analyse static entangling surfaces in asymptotically locally AdS$_6$ spacetimes; the main result of this section is the explicit form of the renormalized area in terms of finite conformal invariants \eqref{eq:A6}. Details of the asymptotic analysis are contained within the appendix. In section \ref{sec:four} we express the renormalized entanglement entropy for spherical entangling regions in terms of the Euler invariant and show that linearized variations can be expressed in terms of the conformal invariant that is linear in the Weyl tensor. We conclude in section \ref{sec:five}. 
	
	\section{Asymptotically AdS$_4$} \label{sec:two}
	
Consider a codimension two static minimal surface $\Sigma$ with boundary $\partial \Sigma$ in an asymptotically locally AdS$_4$ spacetime. The renormalized entanglement entropy $S(\Sigma)$ is expressed in terms of the renormalized area  ${\cal A}(\Sigma)$ as
\begin{equation}
S (\Sigma) = \frac{ {\cal A} (\Sigma)}{4 G_4} \label{area2}
\end{equation}
where $G_4$ is the four-dimensional Newton constant. The renormalized area is \cite{Taylor:2016aoi}
\begin{equation}
{\cal A} ( \Sigma ) = \int_{\Sigma} d^2 x \sqrt{g} - \int_{\partial \Sigma} dx \sqrt{h}.
\end{equation}
Here $g$ is the metric on the minimal surface and $h$ is the metric at the boundary of the minimal surface.

It was shown in \cite{Alexakis:2010zz} that the renormalized area can be expressed in terms of the Euler characteristic of the surface and an integral of local invariants. The analysis of \cite{Alexakis:2010zz} was for two dimensional minimal surfaces in $(d+1)$-dimensional asymptotically locally hyperbolic Einstein spaces i.e. Euclidean signature. This analysis demonstrated that 
\begin{equation}
{\cal A}  (\Sigma) = - 2 \pi \chi(\Sigma) - \frac{1}{2} \int_{\Sigma} d^2 x \sqrt{g} | {K}^s |^2 + \int_{\Sigma} d^2 x \sqrt{g} \widetilde{W}_{3434}  \label{Alex}
\end{equation}
where $\widetilde{W}_{3434}$ is the Weyl curvature of the bulk metric evaluated on any orthonormal basis for the tangent space of the entangling surface and the bulk curvature is normalised to satisfy $R_{\mu \nu} = - d G_{\mu \nu}$. Here $K^s_{ij}$ are the components of the second fundamental form; the index $s$ runs over the directions orthogonal to the surface i.e. $s = 1,2$ in the case of a four-dimensional bulk geometry. Note that the minimal condition implies that $K^s$ is trace free. Each term in \eqref{Alex} is individually finite: the integrands in the last two terms fall off sufficiently quickly near the conformal boundary that the integrals do not have divergent contributions \cite{Alexakis:2010zz}. 

In the case of a static Ryu-Takayanagi entangling surface, the extrinsic curvature in the time direction is zero and by tracelessness of the Weyl curvature the renormalized area reduces to 
\begin{equation}
{\cal A}  (\Sigma) = - 2 \pi \chi(\Sigma) - \frac{1}{2} \int_{\Sigma} d^2 x \sqrt{g} | {K} |^2 - \int_{\Sigma} d^2 x \sqrt{g} \widetilde{W}_{1212} \label{simp1}
\end{equation}
where $K_{ij}$ is the extrinsic curvature of the surface along a spatial section and $\widetilde{W}_{1212}$ is the Weyl curvature evaluated on an orthonormal basis for the normal space of $\Sigma$. Writing the Weyl tensor in this way is to match with our higher dimensional result shown in the later section. 

\subsection{Disk entangling region}

Let us now consider the renormalized entanglement entropy in particular contexts. In pure AdS$_4$ the Weyl tensor vanishes and therefore 
\begin{equation}
S (\Sigma) = - \frac{\pi}{2 G_4} \chi (\Sigma) - \frac{1}{8 G_4}  \int_{\Sigma} d^2 x \sqrt{g} | {K} |^2 \label{bound1}
\end{equation}
Consider a single entangling region in the boundary, which is topologically a disk. The corresponding Ryu-Takayanagi surface has the same topology and accordingly its Euler characteristic $\chi (\Sigma) = 1$. The renormalized entanglement entropy for such surfaces therefore satisfies
\begin{equation}
S (\Sigma) \le - \frac{\pi}{2 G_4}
\end{equation}
with equality in the case of $K_{ij} = 0$. Minimal surfaces satisfy $K=0$; surfaces that in addition satisfy $K_{ij} = 0$, i.e. the traceless part of the extrinsic curvature vanishes, are called umbilic. Umbilic surfaces are locally spherical; the normal curvatures in all directions are equal.

In the specific case of a disk entangling region, the entangling surface indeed has zero extrinsic curvature and is umbilic. This can be seen by 
changing from Poincar\'{e} coordinates:
\begin{equation}
ds^2 = \frac{1}{\rho^2} \left ( - dt^2 + d\rho^2 + dr^2 + r^2 d\phi^2 \right )
\end{equation}
to new coordinates adapted to the entangling surface:
\begin{equation}
\rho = R \sin \theta \qquad
r = R \cos \theta
\end{equation}
so that 
\begin{equation}
ds^2 = \frac{1}{R^2 \sin^2 \theta} \left ( -dt^2 + dR^2 + R^2 (d \theta^2 + \cos^2 \theta d \phi^2 ) \right ).
\end{equation}
The induced metric on an entangling surface of constant $t$ and $R$ can thus be written as 
\begin{equation}
ds^2 = \frac{1}{\sin^2 \theta} \left ( d \theta^2 + \cos^2 \theta d \phi^2 \right ),
\end{equation}
which is independent of both $t$ and $R$, demonstrating that the extrinsic curvatures are zero. 

For a disk entangling region ${\cal D}$, the renormalized entropy is thus directly proportional to the Euler characteristic of the entangling surface. 
As discussed in  \cite{Taylor:2016aoi, Taylor:2016kic}, the renormalized entropy is also related to the F quantity of the corresponding 3d CFT and hence
\begin{equation}
F = - S( {\cal D}) = \frac{\pi}{2 G_4} \chi ( {\cal D} )
\end{equation} 
and the representation of the entanglement entropy in terms of a topological invariant emphasises that this quantity does not depend on any choice of renormalization scheme. 

\bigskip

Now let us consider linearized perturbations around the disk entangling surface in AdS$_4$. Linear and quadratic perturbations around generic minimal surfaces in asymptotically hyperbolic manifolds were discussed in detail in \cite{Alexakis:2010zz}. The analysis of \cite{Alexakis:2010zz} however simplifies considerably for perturbations around the disk entangling surface as both the Weyl and extrinsic curvatures vanish at leading order. Accordingly the only term in the linearized variation is
\be
\delta S = \frac{-1}{4 G_4} \int d^2x \sqrt{g} \; \delta \widetilde{W}_{1212}
\ee
In a subsequent work \cite{FirstLaw} we will show how $\delta {\tilde W}_{1212}$ can be related to the renormalized stress tensor constructed in \cite{deHaro2001} and hence to the variation in the energy; this gives a generalized proof of the first law \cite{Faulkner:2013ica}. 
 
\subsection{Strip entangling region}

Consider now a strip entangling region ${\cal S}$ in pure AdS$_4$. Using the following Poincar\'{e} coordinates
\be
ds^2 = \frac{1}{\rho^2} \left ( -dt^2 + d \rho^2 + dx^2 + dy^2 \right ),
\ee
the entangling surface for a strip entangling region along the $y$ direction can be expressed as 
\be
\frac{d \rho}{dx} = \mp\frac{ \sqrt{\rho_c^4 - \rho^4}}{\rho^2}\label{eq:min4strip}
\ee
where $\rho_c$ is the turning point of the surface and $-$ for $0\leq x\leq\frac{L_x}{2}$ and $+$ for $-\frac{L_x}{2}\leq x\leq 0$.  The width of the strip $L_x$ along the $x$ direction is related to $\rho_c$ as
\be
L_x = 2 \int_{0}^{\rho_c} \frac{\rho^2}{\sqrt{\rho_c^4 - \rho^4}} d \rho = 2 \sqrt{\pi} \frac{\Gamma ( \frac{3}{4} )}{\Gamma  ( \frac{1}{4})} \rho_c. 
\ee
Here implicitly we assume that $L_x \ll L_y$, where $L_y$ is the length of the strip, so that contributions from the corners and short sides 
are negligible. The renormalized area ${\cal A}( {\cal S})$ is then given by 
\be
{\cal A} ( {\cal S}) = - \frac{2 L_y}{\rho_c} \sqrt{\pi} \frac{\Gamma ( \frac{3}{4} )}{\Gamma  ( \frac{1}{4})} = - \frac{L_y L_x}{ \rho_{c}^2}.\label{eq:RAS}
\ee
Since for large $L_y$ the Euler characteristic is negligible and in the limit of the infinite strip $\chi ( {\cal S}) = 0$, and the Weyl curvature vanishes for pure AdS, the renormalized area \eqref{simp1} is given in terms of the integral of the extrinsic curvature over the surface. 

Using $(\ref{eq:min4strip})$ we can pullback the $AdS_4$ metric onto $\mathcal{S}$ to give:
\begin{align}
ds^2=\frac{1}{\rho(x)^2}\left(\frac{\rho_c^4}{\rho(x)^4}dx^2+dy^2\right),
\end{align}
where implicitly $\rho$ is expressed in terms of $x$. The push forward of the unit spatial normal vector is
\begin{align}
n_2=\frac{\rho^3}{\rho^2_c}\left(\frac{\partial}{\partial\rho}\pm\frac{\sqrt{\rho^4_c-\rho^4}}{\rho^4}\frac{\partial}{\partial x}\right).
\end{align}
The interpretation of the two signs is as follows. Let the strip extend from $x = - \frac{1}{2} L_x$ to $x = \frac{1}{2} L_x$. For $x > 0$, the normal to the entangling surface points in the direction of increasing $x$ (positive sign) while for $x < 0$ the normal points in the direction of decreasing $x$ (negative sign). 
Accordingly the induced metric can be written as 
\begin{align}
G^{\mathcal{S}}_{\mu\nu}dx^{\mu}dx^{\nu}&=\left(G_{\mu\nu} - {n_2}_{\mu}{n_2}_{\nu} \right)dx^{\mu}dx^{\nu}\\
&=\frac{1}{\rho^2}\left(\frac{\rho^4_c-\rho^4}{\rho^4_c}d\rho^2-2\frac{\rho^2\sqrt{\rho^4_c-\rho^4}}{\rho^4_c}d\rho dx+\frac{\rho^4}{\rho^4_c}dx^2+dy^2  \right). \nonumber 
\end{align}
The temporal extrinsic curvature vanishes and the spatial extrinsic curvature is given by
\begin{align}
K_{\mu\nu}dx^{\mu}dx^{\nu}=\frac{\rho^4_c-\rho^4}{\rho^4_c}d\rho^2\mp 2\frac{\rho^2\sqrt{\rho^4_c-\rho^4}}{\rho^6_c}d\rho dx+\frac{\rho^4}{\rho^6_c}dx^2-\frac{1}{\rho^2_c}dy^2, \label{eq:KStrip}
\end{align}
The trace of the extrinsic curvature can be easily read off and satisfies the required minimality condition, $K=0$. From $(\ref{simp1})$, the only non vanishing term of the renormalized area is
\begin{align}
\mathcal{A}(\mathcal{S})&=-\frac{1}{2}\int_{\mathcal{S}}d^2x\sqrt{g}K^{\mu\nu}K_{\mu\nu}
=-\frac{1}{2}\int^{\frac{L_y}{2}}_{-\frac{L_y}{2}}dy\int^{\frac{L_x}{2}}_{-\frac{L_x}{2}}dx\frac{\rho^2_c}{\rho^4}\left(\frac{2\rho^4}{\rho^4_c}\right) =-\frac{L_yL_x}{\rho_c^2}.
\end{align}
Note that $K^{\mu\nu}K_{\mu\nu}$ takes the same value for either sign in \eqref{eq:KStrip}.
This matches with the explicit result for the renormalized area of the minimal surface extends from the strip entangling region in $(\ref{eq:RAS})$.

\section{Asymptotically AdS$_6$} \label{sec:three}

Consider a codimension two static minimal surface $\Sigma$ with boundary $\partial \Sigma$ in an asymptotically locally AdS$_6$ spacetime. The renormalized entanglement entropy $S(\Sigma)$ is expressed in terms of the renormalized area  ${\cal A}(\Sigma)$ as
\begin{equation}
S (\Sigma) = \frac{ {\cal A} (\Sigma)}{4 G_6}
\end{equation}
where $G_6$ is the six-dimensional Newton constant. The renormalized area is \cite{Taylor:2016aoi}
\begin{eqnarray}
{\cal A}(\Sigma)  &=& \int_{\Sigma} d^4 x \sqrt{g} - \frac{1}{3} \int_{\partial \Sigma} d^3 x \sqrt{h} \label{renorma} \\
&& - \frac{1}{9} \int_{\partial \Sigma} d^3 x \sqrt{h} \left ( \hat{R}_{aa} - \frac{1}{2} k^2 - \frac{5}{8} \hat{R} \right ). \nonumber
\end{eqnarray}	
Here $g$ is the metric on the minimal surface and $h$ is the metric at the boundary of the minimal surface. $\hat{R}_{aa}$ is the curvature of the metric on the 
boundary of the asymptotically locally AdS$_6$ spacetime, projected to the subspace orthogonal to $\partial \Sigma$. $\hat{R}$ is the Ricci scalar of the boundary curvature and $k^2$ is the square of the extrinsic curvature of $\partial \Sigma$ embedded into $\partial M$, the boundary of the asymptotically locally AdS$_6$ spacetime $M$. The counterterms are sufficient for bulk dimension less than or equal to six; additional divergences arise in higher dimensions \cite{Taylor:2016aoi}. 

\bigskip

Using the Chern-Gauss-Bonnet theorem, the Euler invariant for a four-dimensional manifold with boundary consists of a bulk contribution
\begin{equation}
\chi( \Sigma) = \frac{1}{32 \pi^2} \int_{\Sigma} d^4 x \sqrt{g} \left ( {\cal R}^{ijkl} {\cal R}_{ijkl} - 4 {\cal R}_{ij} {\cal R}^{ij} + {\cal R}^2 \right) \label{euler} 
\end{equation}
(where ${\cal R}$ refers to the intrinsic curvature of the manifold) with
boundary contributions that may be expressed as in \cite{Dowker:1989ue}:
\begin{align}
+\frac{1}{4 \pi^2} \int_{\partial \Sigma} d^3x \sqrt{h} \bigg( & {\cal R}_{ijkl} {\cal K}^{ik} n^{j} n^k - {\cal R}^{ij} {\cal K}_{ij} - {\cal K} {\cal R}_{ij} n^i n^j + \frac{1}{2} {\cal K} {\cal R}\label{euler2} \\
&+ \frac{1}{3} {\cal K}^3  - {\cal K} {\rm Tr} ( {\cal K} ^2) +  \frac{2}{3} {\rm Tr} ( {\cal K}^3) \bigg) \nonumber.
\end{align}
The above formulae used different sign convention to \cite{Dowker:1989ue} and are further explained in the appendix.
Note that this form for the boundary contributions was derived in the context of analysing conformal anomalies on manifolds with boundary. 

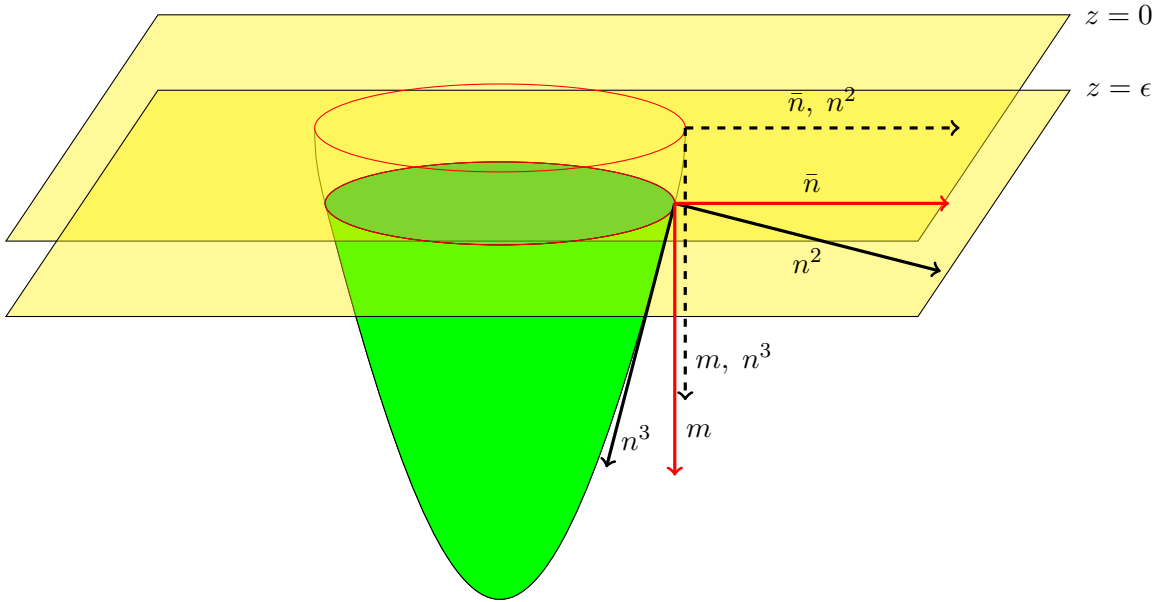
\begin{figure}
\begin{tikzpicture}



\draw[black] (-2.3,0) ..controls (-0.5,-7) and (0.5,-7)..(2.3,0);

\draw[black] (-2.438,1) to [out=-90,in=104.42] (-2.3,0);

\draw[black] (2.438,1) to [out=-90,in=75.58] (2.3,0);

\filldraw[fill=green] (-2.3,0) ..controls (-0.5,-7) and (0.5,-7)..(2.3,0)arc (0:-180:2.3 and 0.55);
\draw[red] (2.3,0)arc (0:-180:2.3 and 0.55);

\filldraw[fill=yellow,fill opacity=0.4] (-6.5,-0.5) -- (5.5,-0.5)-- (7.5,2.5)--(-4.5,2.5)-- cycle;

\filldraw[fill=yellow,fill opacity=0.4] (-6.5,-1.5) -- (5.5,-1.5)-- (7.5,1.5)--(-4.5,1.5)-- cycle;

\draw (7.5,2.5)node{\;\;\;\;\;\;\;\;\;\;\;\;$z=0$};

\draw (7.5,1.5)node{\;\;\;\;\;\;\;\;\;\;\;\;$z=\epsilon$};
\filldraw[fill=green!70!black,fill opacity=0.5] (0,0) ellipse (2.3cm and 0.55cm);

\draw[red] (0,1) ellipse (2.438cm and 0.583cm);

\draw[red] (0,0) ellipse (2.3cm and 0.55cm);

\draw[dashed,->,very thick] (2.438,1)-- node[near end,below](K2){\;\;\;\;\;\;\;\;\;\;\;\;$m,\;n^3$}(2.438,-2.61);

\draw[dashed,->,very thick] (2.438,1)--node[midway,above](k){$\bar{n},\;n^2$}(6.048,1);

\draw[->,very thick] (2.3,0)-- node[pos=0.8,below](K2){\;\;\;\;$n^3$}(1.4,-7/2);
\draw[->,very thick] (2.3,0)--node[midway,below](K3){$n^2$}(5.8,-0.9);
\draw[red,->,very thick] (2.3,0)--node[midway,above](k){\textcolor{black}{$\bar{n}$}}(5.91,0);
\draw[red,->,very thick] (2.3,0)--node[pos=0.9,above](kperp){\;\;\;\;\;\;\textcolor{black}{$m$}}(2.3,-3.61);

\end{tikzpicture}
\captionof{figure}{In this diagram the temporal direction, $n^1$, is suppressed. We can identify two distinct sets of normal directions: $n^2$ is the normal of the $\Sigma$ and $n^3$ is the normal of $\partial\Sigma$ within $\Sigma$, while $\bar{n}$ is the normal of $\partial\Sigma$ on $\partial M$ and $m$ is the normal of $\partial M$ in $M$. On the regulated boundary $\partial M\vert_{z=\epsilon}$, $n^2$ and $n^3$ are not equal to $\bar{n}$ and $m$ respectively. However the normal space is manifestly spanned by both $\{n^s, s=1,2,3\}$ and $\{n^1,\bar{n},m \}$.}
\label{plot:fig1}
\end{figure}

By construction both functionals \eqref{renorma} and \eqref{euler} are finite. However, there are clear conceptual differences between the boundary terms. In the case of the renormalised area, the boundary terms are counterterms, expressed covariantly in terms of Dirichlet data at the conformal boundary. This implies that the boundary terms have to be expressed only in terms of the intrinsic curvature of the conformal boundary, and the extrinsic curvature of the boundary of the entangling surface, embedded into the conformal boundary. 

By contrast, the Euler invariant is expressed entirely in terms of quantities that are intrinsic to the entangling surface itself, with no reference to the embedding of the surface into the six-dimensional bulk manifold. Here the boundary terms are not expressed in terms of Dirichlet data at the boundary of the surface, but involve the extrinsic curvature of the boundary. 

The goal of this section is to relate the renormalised area to the Euler invariant, through the use of Gauss-Codazzi relations and asymptotic analysis. 
Related analysis was carried out in the mathematics literature in \cite{Zhang:2017lcd,Graham:2017bew}  but these works did not use explicit counterterms to define the renormalized area. 

\subsection{Geometric preliminaries}

The extrinsic curvatures of the entangling surface $\Sigma$ are defined by
\begin{align}
K^{s}_{\mu \nu} = g_{\mu}^{\rho}g_{\nu}^{\sigma}\nabla_{\rho}n^s_{\sigma} \label{norm-Sigma}
\end{align}
where the normals to $\Sigma$ are $n^s$ with $s=1,2$; we will denote by $n^1$ the normal in the time direction. 
Similarly, the extrinsic curvature of the boundary entangling surface $\partial\Sigma$ embedded into $\Sigma$ is defined by
\begin{align}
\mathcal{K}_{ij}=-h_{i}^{k}h_{j}^{l}\nabla_{k}n^3_{l}
\end{align}
where $n^3$ is the associated inward pointing normal, as shown in Figure~\ref{plot:fig1}. 

We can define a second set of normals to $\partial \Sigma$, 
$(n^1, \bar{n}, m)$,  where $\bar{n}$ is the normal of $\partial\Sigma$ lying within $\partial M$ and $m$ is the normal of $\partial M$ in $M$.
The extrinsic curvatures corresponding to this second set of normals are defined as
\begin{align}
k_{ij}&=h_{i}^{k}h_{j}^{l}\nabla_{k}\bar{n}_{l} & k^{\perp}_{ij}&=h_{i}^{k}h_{j}^{l}\nabla_{k}m_{l}
\end{align}
The two sets of normal vectors $(n_1,n_2,n_3)$ and $(n_1,\bar{n},m)$ can be related by coordinate transformations.
\begin{align}
n^2&=A\bar{n}+A^{\perp} m\label{eq:NDecomp}\\
n^3&=-A^{\perp}\bar{n}+A m \nonumber
\end{align}
where $A^2+{A^{\perp}}^2=1$ and $A,A^{\perp}\in \mathbb{R}$. The induced metric on $\partial\Sigma$ can also be obtained from either set of normal
\begin{equation}
h_{\mu\nu}=G_{\mu\nu}+n^1_{\mu}n^1_{\nu}-n^2_{\mu}n^2_{\nu}-n^3_{\mu}n^3_{\nu}
=G_{\mu\nu}+n^1_{\mu}n^1_{\nu}-\bar{n}_{\mu}\bar{n}_{\nu}-m_{\mu}m_{\nu.}
\end{equation}
Note that it is often convenient to work in an orthonormal basis, $G_{\mu\nu}e^{\;\mu}_Me^{\;\nu}_N=\eta_{MN}$, such that the 
induced metric on $\Sigma$ is $e^Ae^A=n^3n^3+e^ae^a=g_{\mu\nu}dx^{\mu}dx^{\nu}$ and on $\partial\Sigma$ is $e^ae^a=h_{ij}dx^idx^j$.

\bigskip

As in \cite{Alexakis:2010zz}, we will work in Fefferman-Graham coordinate systems for asymptotically locally AdS metrics:
\begin{align}
ds^2=G_{\mu\nu}dx^{\mu}dx^{\nu} =\frac{dz^2}{z^2}+\gamma_{\alpha\beta}dx^{\alpha}dx^{\beta}
\end{align}
and the metric $\gamma$ admits the expansion
\begin{align}
\gamma_{\alpha\beta}=z^{-2} \left ( \bar{\gamma}_{\alpha\beta}^{(0)}+z^2\bar{\gamma}_{\alpha\beta}^{(2)}+\cdots \right )
\end{align}
Note that this implicitly assumes that the entangling surface is contained within the Fefferman-Graham coordinate patch. 

For static manifolds $\Sigma$ and $\partial\Sigma$ in AdS we can then express the normals as 
\begin{align}
n^1=\frac{dt}{z}, \;\;\;\;m=\frac{dz}{z},\;\;\;\;\bar{n}=\frac{\bar{\alpha} d \bar{f}}{z}.
\end{align} 
(For non-static surfaces one would need to parameterise the timelike normal as $n^1=\frac{\alpha_{\tau}d{\bar{\tau}}}{z}$.) 
The corresponding normal vector fields are
\begin{align}
n_1=e_1=z\frac{\partial}{\partial t}, \;\;\;\;e_m=z\frac{\partial}{\partial z},\;\;\;\;e_{\bar{n}}=\frac{z}{\bar{\alpha}}\frac{\partial}{\partial \bar{f}}
\end{align}
where $\bar{\alpha}$ is a function of $(z,\bar{f})$ only. Using these relations one can decompose bulk curvatures into quantities that are intrinsic and extrinsic to the surface. For example, 
the Lie bracket for the normal vector fields has structure constant $F_{MN}^P$ such that $[e_M,e_N]=F^P_{MN}e_P$. Only the following components are non-vanishing
\begin{align}
F_{m1}^1=-F_{1m}^1=1,\;\;\;\; F_{m\bar{n}}^{\bar{n}}=-F_{\bar{n}m}^{\bar{n}}=1-\bar{\beta}
\end{align} 
where $\bar{\beta}=\frac{z\partial_z\bar{\alpha}}{\bar{\alpha}}$. From these expressions we can then work out the connections and curvature tensors in terms of quantities defined on $\Sigma$. 

\subsection{Gauss-Codazzi relations}	
	
In this section we collect together identities relating the bulk curvature with the intrinsic and and extrinsic curvatures of the entangling surface. First let us note the following relation for the bulk curvature: since the manifold is Einstein with negative cosmological constant, we can express the Riemann curvature in terms of the Weyl curvature as 
 \begin{equation}
 W_{\mu \nu \rho \sigma} = R_{\mu \nu \rho \sigma} + G_{\mu\rho} G_{\nu \sigma} - G_{\mu \sigma} G_{\nu \rho}
 \end{equation}	
where $G_{\mu \nu}$ is the metric on ${\cal M}$. In particular, the Weyl curvature vanishes for anti-de Sitter spacetime itself. 

In this section we will implement Gauss-Codazzi relations for the codimension two surface, taking into account both
(unit) normals to the entangling surface by $n^{s}_{\mu}$ with $s =1,2$. In the context of Ryu-Takayanagi surfaces the extrinsic curvatures in time directions are trivial, but the analysis carried out in this section is more general and does not pick out a distinguished coordinate system for the normal 
directions.  

The Gauss-Codazzi relations then state that:
\begin{equation}
g_{\mu}^{\kappa} g_{\nu}^{\lambda} g_{\rho}^{\tau} g_{\sigma}^{\eta} \; R_{\kappa \lambda \tau \eta} = 
{\cal R}_{\mu \nu \rho \sigma} + \sum_{s=1}^2 (-1)^{s}( K^{s}_{\mu \sigma} K^{s}_{\nu \rho} - K^s_{\mu \rho} K^{s}_{\nu \sigma} )
\end{equation}
where the extrinsic curvatures are defined above in \eqref{norm-Sigma}. (Note that it is often convenient to choose adapted coordinates for the hypersurface.)

The pullback of the bulk curvature can be expressed as
\begin{equation}
g_{\mu}^{\kappa} g_{\nu}^{\lambda} g_{\rho}^{\tau} g_{\sigma}^{\eta} \; R_{\kappa \lambda \tau \eta} = 
g_{\mu}^{\kappa} g_{\nu}^{\lambda} g_{\rho}^{\tau} g_{\sigma}^{\eta} \; W_{\kappa \lambda \tau \eta} + g_{\mu \sigma} g_{\nu \rho} -
g_{\mu\rho} g_{\nu \sigma},
\end{equation}
using $g^{\mu \nu} n^s_{\mu} = 0$. In what follows it is convenient to use a compressed notation to denote the pulled back Weyl curvature as
\begin{equation}
\widetilde{W}_{\mu \nu \rho \sigma} \equiv g_{\mu}^{\kappa} g_{\nu}^{\lambda} g_{\rho}^{\tau} g_{\sigma}^{\eta} \; W_{\kappa \lambda \tau \eta}. 
\end{equation}
Contraction of the Gauss-Codazzi relations gives 
\begin{equation} 
g_{\mu}^{\kappa} g_{\nu}^{\lambda} R_{\kappa \lambda} + g_{\mu}^{\kappa} g_{\nu}^{\lambda} R_{\kappa \tau \lambda \eta} \sum_{s=1}^{2} (-1)^{s-1}  n_s^{\tau} n_s^{\eta} = {\cal R}_{\mu \nu} + \sum_{s=1}^{2} (-1)^{s} K^{s}_{\mu \rho} {K^s}_{\nu}^{\; \rho},	
\end{equation}
where here and in the rest of this sectoin we show the normal index $s$ as a subscript to improve the clarity of equations. Contracting further gives
\begin{equation} 
R +  2g_{\mu}^{\kappa} g_{\nu}^{\lambda} R_{\kappa\lambda} \sum_{s=1}^{2} (-1)^{s-1}n_s^{\mu} n_s^{\nu}-2R_{\mu\nu\rho\sigma} n_1^{\mu} n_2^{\nu} n_1^{\rho} n_2^{\sigma} = {\cal R} + \sum_{s=1}^{2} (-1)^{s} K^{s}_{\mu \rho} {K^s}^{\;\mu\rho}	
\end{equation}
where we use the fact that the surface is minimal so $K^s = 0$. In our case, the background manifold is Einstein, for which the Ricci curvature can conveniently be normalised as 
\begin{equation}
R_{\mu \nu} = - d G_{\mu \nu}
\end{equation}
for asymptotically locally AdS$_{(d+1)}$ spacetimes. Using the fact that $g_{\mu \nu} n_s^{\n} = 0$, we can thus write
\begin{equation}
{\cal R}_{\mu \nu} + \sum_{s=1}^{2} (-1)^{s} K^{s}_{\mu \rho} {K^s}_{\nu}^{\; \rho}	= - (d-2) g_{\mu \nu}- g^{\lambda}_{\mu}g^{\tau}_{\nu}W_{\lambda\rho\tau\sigma} \sum_{s=1}^{2} (-1)^{s}  n_s^{\rho} n_s^{\sigma},
\end{equation}
and
\begin{equation}
{\cal R} + \sum_{s=1}^{2} (-1)^{s} K^{s}_{\mu \nu} {K^s}^{\mu \nu}	= -(d-2)(d-1)-2W_{\mu\nu\rho\sigma} n_1^{\mu} n_2^{\nu} n_1^{\rho} n_2^{\sigma},
\end{equation}
where we use the fact that the dimension of the entangling surface is $(d-1)$. 

For notational convenience we will define the combinations
\begin{equation}
H_{\mu\nu\rho\sigma}=\sum_{s=1}^2 (-1)^{s} (K^{s}_{\mu \sigma} K^{s}_{\nu \rho} - K^s_{\mu \rho} K^{s}_{\nu \sigma} ), 
\end{equation}
as well as 
\begin{equation}
\widetilde{W}_{\mu n \nu n}=g^{\lambda}_{\mu}g^{\tau}_{\nu}W_{\lambda\rho\tau\sigma} \sum_{s=1}^{2} (-1)^{s}  n_s^{\mu} n_s^{\nu}
\end{equation}
and 
\begin{equation}
\widetilde{W}_{1212}=W_{\mu\nu\rho\sigma} n_1^{\mu} n_2^{\nu} n_1^{\rho} n_2^{\sigma}.
\end{equation}
The Gauss-Codazzi relations can then be used to rewrite the bulk term in the Euler invariant as follows. The Riemann curvature terms give
\begin{eqnarray}
{\cal R}_{\mu \nu \rho \sigma} {\cal R}^{\mu \nu \rho \sigma} &=& 2 (d-1) (d-2)+4H+ H_{\mu\nu\rho\sigma}H^{\mu\nu\rho\sigma} \\ 
&& -4 \widetilde{W}+\widetilde{W}_{\mu\nu\rho\sigma}\widetilde{W}^{\mu\nu\rho\sigma}-2\widetilde{W}_{\mu\nu\rho\sigma} H^{\mu\nu\rho\sigma}.\nonumber
\end{eqnarray}
The Ricci curvature terms give
\begin{eqnarray}
{\cal R}_{\mu \nu} {\cal R}^{\mu \nu} &=& (d-2)^2(d-1) + 2(d-2)H + H_{\mu\nu}H^{\mu\nu}  \\
&& +2(d-2)\widetilde{W}_{nn}+ \widetilde{W}_{\mu n \nu n}\widetilde{W}^{\mu n \nu n}+ 2\widetilde{W}_{\mu n \nu n}H^{\mu\nu}\nonumber
\end{eqnarray}
while the Ricci scalar terms give
\begin{eqnarray}
{\cal R}^2 &=& (d-2)^2(d-1)^2 + 2(d-2)(d-1)H + H^2  \\
&&+ 4(d-2)(d-1)\widetilde{W}_{1212} + 4\widetilde{W}_{1212}^2+ 4H\widetilde{W}_{1212}. \nonumber
\end{eqnarray}
Here $H$ and $H_{\mu \nu}$ can be expressed as
\be
H_{\mu \nu} = \sum_{s=1}^2 (-1)^{s}  K^{s}_{\mu \sigma} {K^{s}}_{\nu}^{\; \; \sigma} \qquad
H =  \sum_{s=1}^2 (-1)^{s} K^{s}_{\mu \nu} {K^{s}}^{\mu \nu}.
\ee
Combining these terms for the case of $d=5$ (AdS$_6$), we obtain an expression for the Euler invariant of the form:
\begin{equation}
\chi (\Sigma) = \frac{1}{32 \pi^2} \int_{\Sigma} d^4 x \sqrt{g} \left ( 24 + \Delta \chi (K^s, \widetilde{W} ) \right )+\frac{1}{4\pi^2}\int_{\partial\Sigma}d^3x \sqrt{h}\partial E_4 \label{Chi-exp}
\end{equation}	
where the functional appearing in the volume term takes the form
\begin{align}
\Delta \chi =& 4H  + 8 \widetilde{W}_{1212}
+ H^2 -4H_{\mu\nu}H^{\mu\nu} + H_{\mu\nu\rho\sigma}H^{\mu\nu\rho\sigma} \label{DeltaChi}\\
& +4\widetilde{W}_{1212}^2-4\widetilde{W}_{\mu n \nu n}\widetilde{W}^{\mu n \nu n}+\widetilde{W}_{\mu \nu\rho\sigma}\widetilde{W}^{\mu \nu\rho\sigma}\nonumber\\
& +4H\widetilde{W}_{1212}-8H^{\mu\nu}\widetilde{W}_{\mu n \nu n}-2H^{\mu \nu\rho\sigma}\widetilde{W}_{\mu \nu\rho\sigma}\nonumber
\end{align}
The notation chosen for the volume term reflects the fact that $\Delta \chi$ vanishes for spherical entangling surfaces ($K^s_{\mu \nu} = 0$) in pure AdS ($W_{\mu \nu \rho \sigma}$), as we discuss in the next section. 	

To simplify this expression we have used the following expressions for the projected and contracted Weyl tensor
\begin{align}
&\widetilde{W}=\widetilde{W}^{AB}_{\;\;\;\;AB}=W^{AB}_{\;\;\;\;AB} & \widetilde{W}_{22}&=W_{\;\;2A2}^A \label{eq:Weyls} \\
&\widetilde{W}_{1212}=W_{1212} & \widetilde{W}_{11}&=W_{\;\;1A1}^A. \nonumber
\end{align}
Since Weyl tensor is traceless, we can write the curvatures in $(\ref{eq:Weyls})$ in terms of each other as
\begin{align}
W^{AB}_{\;\;\;\;AB}&=W^{MN}_{\;\;\;\;MN}-2\left(W^{A1}_{\;\;\;\;A1}+W^{A2}_{\;\;\;\;A2}+W^{12}_{\;\;\;\;12}\right) \\
\widetilde{W}&=2\widetilde{W}_{11}-2\widetilde{W}_{22}+2\widetilde{W}_{1212} \nonumber 
\end{align}
%
and thus
\begin{equation}
\widetilde{W}=-\widetilde{W}_{nn}=-2\widetilde{W}_{1212}
\end{equation}
Therefore the contributions linear in the Weyl tensor can be written in terms of the projection of the Weyl tensor onto $N\Sigma$, $\widetilde{W}_{1212}$.

\bigskip

We now need to express the boundary contributions to the Euler density integral in terms of extrinsic curvatures and the Weyl tensor.
We first define the extrinsic curvature, $\mathscr{K}$, of $\partial M$ embedded into $M$,
\begin{align}
\mathscr{K}_{\mu\nu}=(\delta^{\rho}_{\mu}-m^{\rho}_{\mu})(\delta^{\sigma}_{\nu}-m^{\sigma}_{\nu})\nabla_{\rho}m_{\sigma}.
\end{align}
Using the definition of the extrinsic curvature of $\partial\Sigma$ pointing out of the boundary $k^{\perp}_{\mu\nu}\vcentcolon = h^{\rho}_{\mu}h^{\sigma}_{\nu}\nabla_{\rho}m_{\sigma}$, one can show that
\begin{align}
\mathscr{K}_{\mu\nu}=k^{\perp}_{\mu\nu}-(1-\bar{\beta})\bar{n}_{\mu}\bar{n}_{\nu}+n^1_{\mu}n^1_{\nu}.\label{eq:kz}
\end{align}
The trace of $\mathscr{K}$ is
\begin{align}
\mathscr{K}=k^{\perp}-2+\bar{\beta} \label{eq:Tkz}
\end{align}
and the trace of the product is
\begin{align}
\mathscr{K}_{\mu\nu}\mathscr{K}^{\mu\nu}=k^{\perp}_{\mu\nu}k^{\perp\;\mu\nu}+2-2\bar{\beta}+\bar{\beta}^2\label{eq:Tkzkz}
\end{align}
The intrinsic curvature of the boundary $\partial M$, $\hat{R}_{\mu\nu\rho\sigma}$, is related to the projection and contraction of Weyl tensor and $\mathscr{K}$ by the Gauss-Codazzi equation, giving the following relations
\begin{align}
&\hat{R}_{1\bar{n}1\bar{n}}=1+ W_{1\bar{n}1\bar{n}}-\mathscr{K}_{1\bar{n}}\mathscr{K}_{\bar{n}1}+\mathscr{K}_{11}\mathscr{K}_{\bar{n}\bar{n}} \nonumber \\
&\hat{R}_{11}=d-1-W_{1m1m}-\mathscr{K}_{1}^{\;\rho}\mathscr{K}_{\rho1}+\mathscr{K}_{11}\mathscr{K} \nonumber \\
&\hat{R}_{\bar{n}\bar{n}}=-d+1-W_{\bar{n}m\bar{n}m}-\mathscr{K}_{\bar{n}}^{\;\rho}\mathscr{K}_{\rho\bar{n}}+\mathscr{K}_{\bar{n}\bar{n}}\mathscr{K}
\nonumber \\
&\hat{R}=-d(d-1)-\mathscr{K}_{\mu\nu}\mathscr{K}^{\mu\nu}+\mathscr{K}^2  \\
&\hat{R}_{ij}=-(d-1)h_{ij}-W_{imjm}-\mathscr{K}_{i}^{\;\rho}\mathscr{K}_{\rho j}+\mathscr{K}_{ij}\mathscr{K} \nonumber \\
&\hat{R}_{i\bar{n}j\bar{n}}=-h_{ij}+W_{i\bar{n}j\bar{n}}-\mathscr{K}_{i\bar{n}}\mathscr{K}_{\bar{n}j}+\mathscr{K}_{ij}\mathscr{K}_{\bar{n}\bar{n}} \nonumber \\
&\hat{R}_{i1j1}=h_{ij}+W_{i1j1}-\mathscr{K}_{i1}\mathscr{K}_{1j}+\mathscr{K}_{ij}\mathscr{K}_{11}. \nonumber
\end{align}
Substituting $\mathscr{K}$ terms using  $(\ref{eq:kz})$, $(\ref{eq:Tkz})$ and $(\ref{eq:Tkzkz})$ we obtain
\begin{align}
&\hat{R}_{1\bar{n}1\bar{n}}= W_{1\bar{n}1\bar{n}}+\bar{\beta} \nonumber \\
&\hat{R}_{11}=d-2+k^{\perp}-W_{1m1m}+\bar{\beta} \nonumber \\
&\hat{R}_{\bar{n}\bar{n}}=-d+2 -k^{\perp}-W_{\bar{n}m\bar{n}m}+(k^{\perp}-1)\bar{\beta} \nonumber \\
&\hat{R}=-d(d-1)+2-4k^{\perp}+{k^{\perp}}^2-k^{\perp}_{ij}k^{\perp\;ij}+2(k^{\perp}-1)\bar{\beta}\label{eq:Rhat}\\
&\hat{R}_{ij}=-(d-1)h_{ij}-2k^{\perp}_{ij}-k^{\perp\;k}_ik^{\perp}_{kj}+k^{\perp}_{ij}k^{\perp}-W_{imjm}+k^{\perp}_{ij}\bar{\beta} \nonumber \\
&\hat{R}_{i\bar{n}j\bar{n}}=-h_{ij}-k^{\perp}_{ij}+W_{i\bar{n}j\bar{n}}+k^{\perp}_{ij}\bar{\beta} \nonumber \\
&\hat{R}_{i1j1}=h_{ij}+k^{\perp}_{ij}+W_{i1j1}. \nonumber 
\end{align}

The intrinsic curvature $\mathcal{R}$ terms on the surface given in $(\ref{euler2})$ are related to the curvatures of the boundary of the entangling surface $\partial\Sigma$, $\overline{R}$, by additional Gauss-Codazzi relations:
\begin{align}
\mathcal{K}\left(\frac{1}{2}\mathcal{R}-\mathcal{R}_{\mu\nu}n^{3\;\mu}n^{3\;\nu}\right)=\frac{1}{2}\mathcal{K}\left(\overline{R}-\mathcal{K}^2+\mathcal{K}_{ij}\mathcal{K}^{ij}\right)
\end{align}
and
\begin{align}
-\mathcal{K}^{ij}\mathcal{R}_{\mu i \nu j}(g^{\mu\nu}-n^{3\;\mu}n^{3\;\nu})=-\mathcal{K}^{ij}\left(\overline{R}_{ij}+\mathcal{K}_{ik}\mathcal{K}^k_j-\mathcal{K}_{ij}\mathcal{K}\right).
\end{align}
We can again use Gauss-Codazzi relations to transform relate quantities on $\partial \Sigma$ to quantities in $\partial M$:
\begin{align}
&\overline{R}=\hat{R}+2\hat{R}_{11}-2\hat{R}_{\bar{n}\bar{n}}-2\hat{R}_{1\bar{n}1\bar{n}}+k^2-k_{ij}k^{ij}\\
&\overline{R}_{ij}=\hat{R}_{ij}-\hat{R}_{i\bar{n}j\bar{n}}+\hat{R}_{i1j1}+k_{ij}k-k^{\;k}_ik_{kj}. \nonumber 
\end{align}
Expressing Riemann tensors in terms of Weyl tensors gives
\begin{align}
&\overline{R}=-(d-1)(d-4)+k^2-k_{ij}k^{ij}+k^{\perp}_{ij}k^{\perp\;ij}-2\left(W_{1m1m}+W_{1\bar{n}1\bar{n}}-W_{\bar{n}m\bar{n}m}\right)\\
&\overline{R}_{ij}=-(d-1)h_{ij}+k_{ij}k-k^k_ik_{kj}+k^{\perp}_{ij}k-k^{\perp\;k}_ik^{\perp}_{kj}+W_{i1j1}-W_{i\bar{n}j\bar{n}}-W_{imjm}. \nonumber 
\end{align}
Specialising to $d=5$ these expressions reduce to:
\begin{align}
&\overline{R}=-4+k^2-k_{ij}k^{ij}+k^{\perp}_{ij}k^{\perp\;ij}-2\left(W_{1m1m}+W_{1\bar{n}1\bar{n}}-W_{\bar{n}m\bar{n}m}\right)\\
&\overline{R}_{ij}=-4h_{ij}+k_{ij}k-k^k_ik_{kj}+k^{\perp}_{ij}k-k^{\perp\;k}_ik^{\perp}_{kj}+W_{i1j1}-W_{i\bar{n}j\bar{n}}-W_{imjm}. \nonumber
\end{align}
The decomposition of $\mathcal{K}$ into $k,k^{\perp}$ is straightforward:
\begin{align}
\mathcal{K}_{ij}=+A^{\perp}k_{ij}-Ak^{\perp}_{ij}.
\end{align}
Our final expression for the boundary contributions to the Euler density can be written in terms of projections of the Weyl tensor and extrinsic curvatures of $\partial\Sigma$ tangent and normal to $\partial M$,
\begin{align}
\partial E_4&=-(A^{\perp}k-A k^{\perp})\left(W_{1\bar{n}1\bar{n}}+W_{1m1m}-W_{\bar{n}m\bar{n}m}\right) \nonumber
\\
&+(A^{\perp}k^{ij}-A k^{\perp\;ij})\left(-W_{i1j1}+W_{imjm}+W_{i\bar{n}j\bar{n}} \right)\nonumber\\
&+Ak^{\perp}-\left(\frac{A}{2}-\frac{A^3}{6}\right){k^{\perp}}^3-\left(A-\frac{A^3}{3}\right) k^{\perp}_{ij}k^{\perp\;jk}k^{\perp\;i}_{\;k}-\left(-\frac{3A}{2}+\frac{A^3}{2}\right)k^{\perp}k^{\perp}_{ij}k^{\perp\;ij}\nonumber\\
&-A^{\perp}k-\left(-\frac{A^{\perp}}{2}-\frac{A^2A^{\perp}}{2}\right)k {k^{\perp}}^2-\left(\frac{A^{\perp}}{2}-\frac{A^2A^{\perp}}{2}\right)k k^{\perp}_{ij}k^{\perp\;ij}\nonumber\\
&-\left(-A^{\perp}+A^2A^{\perp}\right)k_{ij}k^{\perp\;jk}k^{\perp \;i}_{\;k}-\left(A^{\perp}-A^2A^{\perp}\right)k^{\perp}k_{ij}k^{\perp\;ij} 
\label{eq:FdE4}\\
&-\left(\frac{A}{2}-\frac{A{A^{\perp}}^2}{2}\right)k^2k^{\perp}-\left(-\frac{A}{2}+\frac{A{A^{\perp}}^2}{2}\right)k_{ij}k^{ij}k^{\perp}\nonumber\\
&-\left(A-A{A^{\perp}}^2\right)k_{ij}k^{jk}k^{\perp \;i}_{\;k}-\left(-A+A{A^{\perp}}^2\right) k k_{ij}k^{\perp\;ij}\nonumber\\
&-\left(-\frac{A^{\perp}}{2}+\frac{{A^{\perp}}^3}{6}\right)k^3-\left(-A^{\perp}+\frac{{A^{\perp}}^3}{3} \right)k_{ij}k^{jk}k^{\;i}_k-\left(\frac{3A^{\perp}}{2}-\frac{{A^{\perp}}^3}{2}\right)k k_{ij}k^{ij}. \nonumber 
\end{align}
We will use this expression in what follows, comparing the boundary terms in the Euler characteristic with those in the renormalized area. 

\subsection{Asymptotic analysis}

The ultimate goal is to express the Euler characteristic as a linear combination of the renormalized area ${\cal A} (\Sigma)$ and other finite contributions i.e. 
\begin{equation}
\chi (\Sigma) = \frac{3}{4 \pi^2} {\cal A} (\Sigma) + \cdots
\end{equation}
where the ellipses denote contributions that are finite term by term. In this section we will show that the finite contributions are such that
\begin{eqnarray}
{\cal A}(\Sigma) &=& \frac{4 \pi^2}{3} \chi (\Sigma) - \frac{1}{6} {\cal H}(\Sigma) - \frac{1}{3} {\cal W}(\Sigma) \label{eq:A6} \\
&& 
- \frac{1}{24}  \int_{\Sigma} d^4 x \sqrt{g} \left ( H^2 - 4 H_{\mu \nu} H^{\mu \nu} +  H_{\mu\nu\rho\sigma}H^{\mu\nu\rho\sigma} \right .  \nonumber \\
&& \left . +4\widetilde{W}_{1212}^2
-4\widetilde{W}_{\mu n \nu n}\widetilde{W}^{\mu n \nu n}+\widetilde{W}_{\mu \nu\rho\sigma}\widetilde{W}^{\mu \nu\rho\sigma} \right ), \nonumber 
\end{eqnarray}
where the finite terms ${\cal H}(\Sigma)$ and ${\cal W}(\Sigma)$ are defined in \eqref{eq:hsigma} and \eqref{eq:wsigma}, respectively,
This formula is the direct generalisation of the corresponding expression for two-dimensional surfaces given in \eqref{Alex}.

\bigskip

To determine the terms arising in this expression, we need to consider the asymptotic analysis of the bulk and boundary terms in the Euler characteristic and the renormalized area. To compare terms between the Euler characteristic and the renormalized area, it is convenient to convert quantities written with respect to quantities intrinsic to $\Sigma$ into quantities expressed with respect to $\partial M$ and $\partial\Sigma$. Intuitively, it is apparent that the extrinsic curvature $K^{(2)}$ of $\Sigma$ and $\mathcal{K}$ of $\partial\Sigma$ can be expressed in terms of the two extrinsic curvatures $k, k^{\perp}$. Indeed, by decomposing the metric and normal of $\Sigma$ into boundary components we can write $K^{(2)}$ as a combination of $k, k^{\perp}$ plus additional terms.  

\bigskip 

The integration in the $M$ is regulated by restricting integration up to the regulated boundary $\partial M_{\epsilon}\vcentcolon = M\vert_{z=\epsilon}$. The regulated divergences in Euler characteristic integral
\begin{align}
\chi ({\Sigma_{\epsilon}}) = \frac{1}{32 \pi^2} \int_{\Sigma_{\epsilon}} d^4 x \sqrt{g} \left ( 24 + \Delta \chi \right )  + \frac{1}{4\pi^2}\int_{\partial\Sigma_\epsilon}d^3x \sqrt{h} \; \partial E_4
\end{align}
come from the bulk terms up to order $z^4$ and boundary terms up to order $z^3$. Clearly by construction all such divergences cancel, as the Euler characteristic is finite, but to compare with the renormalized area we need to identify which terms are finite and which include regulated divergences. 
In what follows we will show that each term in 
\begin{eqnarray}
&& \int_{\Sigma_{\epsilon}} d^4 x \sqrt{g} \left ( H^2 - 4 H_{\mu \nu} H^{\mu \nu} +  H_{\mu\nu\rho\sigma}H^{\mu\nu\rho\sigma}  +4\widetilde{W}_{1212}^2
\right . \\
&& \qquad \qquad \qquad  \left .  -4\widetilde{W}_{\mu n \nu n}\widetilde{W}^{\mu n \nu n}+\widetilde{W}_{\mu \nu\rho\sigma}\widetilde{W}^{\mu \nu\rho\sigma} \right ) \nonumber 
\end{eqnarray}
is individually finite, while the other bulk contributions 
\begin{equation}
\frac{1}{32 \pi^2}   \int_{\Sigma_{\epsilon}} d^4 x \sqrt{g} \left ( 24 + 4 H + 8  \widetilde{W}_{1212} \right ) \label{reg1}
\end{equation}
each have regulated divergences. As we will be comparing regulated divergences of the Euler characteristic with those in the renormalised area, and the latter assumes static embedding, we will set $K^{(1)}_{\mu \nu} = 0$ for the rest of this section.

\bigskip

We need to calculate the asymptotic expansions of the geometric quantities appearing in the Euler characteristic. We begin with the normal vectors defined in \eqref{eq:NDecomp}. If we expand $A^{\perp}$ in $z$ and apply the boundary condition of $A^{\perp}(z=0)=0$ we obtain the leading term in the $z$ power series to be $A^{\perp}=zA^{\perp}_{(0)}+\cdots$. Similarly, the leading term in the $A$ asymptotic series is $A=1+\cdots$. From the relation $A^2+{A^{\perp}}^2=1$ we can thus conclude that the asymptotic expansion for $A$ and $A^{\perp}$ is 
\begin{align}
\begin{split}
A&=1-\frac{1}{2}z^2{A^{\perp}_{(0)}}^2+O(z^4)\\
A^{\perp}&=z{A^{\perp}_{(0)}}+z^3A^{\perp}_{(2)}+O(z^5)
\end{split}\label{eq:AA}
\end{align}
Hence $A,A^{\perp}$ have even and odd power series of $z$ respectively.

The asymptotic analysis for extrinsic curvatures is worked out in the appendix. The trace of the extrinsic curvature behaves as
\begin{align}
K^{(2)} \sim O(z) 
\end{align}
while the trace of the product of extrinsic curvature 
\begin{align}
K^{(2)}_{\mu\nu}K^{(2)\mu\nu} \sim O(z^2).
\end{align}
Accordingly $H$ is of order $z^2$ but terms quadratic in $H$ are of order $z^4$ so do not contribute to the regulated divergences. 
We can write explicit expansions 
\begin{align}
({K^{(2)}})^2 =z^2\left(\bar{k}_{(0)}^2+9{A^{\perp}_{(0)}}^2-6{A^{\perp}_{(0)}}\bar{k}_{(0)} \right)+O(z^4)
\end{align}
and
\begin{align}
K^{(2)}_{ij}K^{(2)ij}&=z^2\left(\bar{k}_{(0) ij}\bar{k}_{(0)}^{ij}+3{A^{\perp}_{(0)}}^2-2A^{\perp}_{(0)} \bar{k}_{(0)} \right)+O(z^6)
\end{align}
where 
\begin{equation}
k =  \bar{k}_{(0)} z + \cdots,
\end{equation}
and
\begin{equation}
k_{ij}k^{ij}\ = z^2 \bar{k}_{(0)ij}\bar{k}_{(0)}^{ij} + \cdots 
\end{equation}
According the regulated divergences from the term linear in $H$ gives 
\begin{align}
\int_{\Sigma_{\epsilon} } d^4x \sqrt{g} H
\rightarrow \int_{\partial\Sigma_{\epsilon}}d^3x\sqrt{h}\;\left ( \epsilon^2\left ( \bar{k}_{(0)ij}\bar{k}_{(0)}^{ij}-\bar{k}_{(0)}^2+4A^{\perp}_{(0)}\bar{k}_{(0)}-6{A^{\perp}_{(0)}}^2 \right )  + \cdots \right )\label{eq:Kdiv}
\end{align}
where the ellipses denote terms that do not contribute in the limit $\epsilon \rightarrow 0$. 

Let us now consider the asymptotic behaviour of the projections and contractions of the Weyl tensors. In our gauge choice $\widetilde{W}, \widetilde{W}_{11}, \widetilde{W}_{22}$ and $\widetilde{W}_{1212}$ are of order $O(z^2)$ and hence only terms linear in the Weyl tensor contribute to the regulated divergences. If the Weyl tensor admits an expansion
\begin{equation}
\widetilde{W}_{1212}=z^2\overline{W}_{1212} + \cdots, 
\end{equation}
leading to regulated divergences
\begin{align}
\int_{\Sigma_{\epsilon} }d^4x  \sqrt{g} \widetilde{W}_{1212} \rightarrow \int_{\partial\Sigma_{\epsilon}}d^3x \sqrt{h}\; \epsilon^2\overline{W}_{1212} \label{eq:Wdiv}
\end{align}
The regulated divergences \eqref{reg1} are obtained from combining $(\ref{eq:Kdiv})$ and $(\ref{eq:Wdiv})$
\begin{eqnarray}
&& \frac{2}{3 \pi^2} \int_{\Sigma_{\epsilon}} d^4x\sqrt{g}\; \label{eq:Ediv} \\
&& +\frac{1}{8 \pi^2} \int_{\partial\Sigma_{\epsilon}} d^3 x\sqrt{h} \epsilon^2 \left (\left[\bar{k}_{(0)ij}\bar{k}_{(0)}^{ij}-\bar{k}_{(0)}^2+4A^{\perp}_{(0)}\bar{k}_{(0)}-6{A^{\perp}_{(0)}}^2\right]+ 2 \overline{W}_{1212}\right ). \nonumber 
\end{eqnarray}
Here we do not explicitly analyse the regulated divergences of the first (area) term, as this was already done in \cite{Taylor:2016aoi}, as we will use below. The expression above can be simplified using the minimal condition for the surface: as explained in the appendix, $K^{(2)} = 0$ implies that
\be
A^{\perp}_{(0)} = \frac{1}{3} \bar{k}_{(0)}
\ee
and therefore $A^{\perp}_{(0)}$ can be eliminated. 

Let us now consider the regulated divergences of the boundary terms in the Euler characteristic. As mentioned in the beginning of the section, only terms of order $O(z^3)$ in $\partial E_4$ contribute to divergences. These terms are analysed in the appendix; the regulated divergences take the form
\begin{align}
 \int_{\partial\Sigma_{\epsilon}}d^3x\sqrt{h}\partial E_4 & \rightarrow 
- \int_{\partial\Sigma_{\epsilon}}d^3x\sqrt{h}\left(1+ \epsilon^2  ( \overline{W}_{1212}- \frac{1}{3}  \bar{k}_{(0)}^2
+ \frac{1}{2} \bar{k}_{(0) ij}\bar{k}^{ij}_{(0)} ) \right).\label{eq:d4div}
\end{align}
By construction the regulated divergences of the boundary terms in the Euler characteristic cancel those from the bulk terms. 

\bigskip

We can now express the regulated contributions in \eqref{eq:Ediv} and \eqref{eq:d4div} in terms of the renormalized area
\begin{align}
 {\cal A }(\Sigma_{\epsilon})= \int_{\Sigma_{\epsilon}} d^4x\sqrt{g} + \frac{1 }{3} \int_{\partial\Sigma_{\epsilon}} d^3 x\sqrt{h}\bigg(-1+\frac{1}{6}k^2\bigg),
\end{align}
and two other integrals that are finite in the limit of $\epsilon \rightarrow 0$:
\begin{equation}
{\cal H}(\Sigma_{\epsilon}) := \int_{\Sigma_{\epsilon}} d^4 x \sqrt{g} H - \int_{\partial {\Sigma_{\epsilon}}} d^3 x \sqrt{h} \left ( k_{ij} k^{ij} - \frac{1}{3} k^2 \right ); \label{eq:hsigma}
\end{equation}
and 
\begin{equation}
{\cal W} (\Sigma_{\epsilon}) := \int_{\Sigma_{\epsilon}} d^4 x \sqrt{g} \widetilde{W}_{1212}  - \int_{\partial{\Sigma_{\epsilon}}} d^3 x \sqrt{h} W_{1212}. \label{eq:wsigma}
\end{equation}
The regulated terms in the Euler characteristic then combine to give
\begin{equation}
\frac{3}{4 \pi^2} {\cal A}(\Sigma_{\epsilon}) + \frac{1}{8\pi^2} {\cal H}(\Sigma_{\epsilon}) + \frac{1}{4 \pi^2} {\cal W}(\Sigma_{\epsilon}),\label{eq:renQ}
\end{equation}
and thus, reinstating the bulk contributions to the Euler characteristic that are individually finite, we obtain the final expression for the renormalized area \eqref{eq:A6}. 

\bigskip

Note that the extra counterterms for the renormalized area integral $(\ref{renorma})$ vanishes in the limit $z\rightarrow 0$. It can be seen from $(\ref{eq:Rhat})$. As $W_{1m1m},W_{\bar{n}m\bar{n}m} \sim O(z^4)$ the individual Ricci terms are
\begin{align}
\begin{split}
&\hat{R}=-8\bar{\beta}+O(z^4)\\
&\hat{R}_{11}=\bar{\beta}+O(z^4)\\
&\hat{R}_{\bar{n}\bar{n}}=-4\bar{\beta}+O(z^4)
\end{split}
\end{align}
Since the definition of the projected Ricci curvature $\hat{R}_{aa}$ is
\begin{align}
\hat{R}_{aa}&:=-\hat{R}_{11}+\hat{R}_{\bar{n}\bar{n}}\\
\hat{R}_{aa}&:=-5\bar{\beta}+O(z^4),\nonumber
\end{align}
the Ricci counterterms $\hat{R}_{aa}-\frac{5}{8}\hat{R}$ is
\begin{align}
-\hat{R}_{11}+\hat{R}_{\bar{n}\bar{n}}-\frac{5}{8}\hat{R}= 0+O(z^4).
\end{align}
The order of this term is great than $z^3$ therefore it vanishes in the boundary integral as $z\rightarrow 0$.

\section{Spherical entangling surface in AdS$_6$ and linear perturbations} \label{sec:four}	

Consider AdS$_6$ written in Poincar\'{e} coordinates as:
\begin{equation}
ds^2 = \frac{1}{\rho^2} \left ( - dt^2 + d\rho^2 + dr^2 + r^2 d \Omega^2 \right )
\end{equation}
We can introduce new coordinates adapted to the entangling surface ${\cal S}$ associated with a spherical entangling region
\begin{equation}
\rho = R \sin \theta \qquad
r = R \cos \theta
\end{equation}
so that 
\begin{equation}
ds^2 = \frac{1}{R^2 \sin^2 \theta} \left ( -dt^2 + dR^2 + R^2 (d \theta^2 + \cos^2 \theta d \Omega^2 ) \right ).
\end{equation}
The induced metric on the entangling surface ${\cal S}$ of constant $t$ and $R$ can thus be written as 
\begin{equation}
ds^2 = \frac{1}{\sin^2 \theta} \left ( d \theta^2 + \cos^2 \theta d \Omega^2 \right ).
\end{equation}
This parameterisation makes manifest that the extrinsic curvatures of ${\cal S}$ within ${\cal M}$ are zero: the induced metric
is independent of the coordinates $t$ and $R$. One can then change coordinates as $u = - \log ( \tan (\theta/2) )$ to write the induced
metric as 
\begin{equation}
ds^2 = du^2 + \sinh^2 u d \Omega^2, 
\end{equation}
i.e. making manifest that the metric on the entangling surface is global AdS with unit radius. 	

\bigskip

The bulk contribution to the Euler invariant is thus
\begin{equation}
\frac{\Omega_3}{32 \pi^2} \left ( 16 + e^{3 \bar{u}} - 9 e^{\bar{u}} + \cdots \right ) 
\end{equation}
where we have regulated the boundary at $u = \bar{u} \gg 1$, and dropped terms that are zero when $\bar{u} \rightarrow \infty$. Here $\Omega_3 = 2 \pi^2$ is the volume of a three sphere of unit radius.  

Calculation of the boundary contributions to the Euler invariant \eqref{euler2} is more complicated. We need the following expressions:
\begin{eqnarray}
{\cal R}_{ijkl}  {\cal K}^{jl} n^i n^k &=& - 3 \frac{ \cosh(u)}{ \sinh (u)}; \qquad 
{\cal K} = 3 \frac{ \cosh(u)}{\sinh (u)}; \\
{\cal R}_{ij} {\cal K}^{ij} &=& - 9 \frac{ \cosh(u)}{\sinh(u)};  \qquad 
{\cal R}_{ij} n^i n^j = -3; \nonumber \\
{\rm Tr} ({\cal K}^2 ) &=& 3 \frac{\cosh^2(u)}{\sinh^2(u)};  \qquad 
{\rm Tr} ( {\cal K}^3 ) = 3 \frac{ \cosh^3(u)}{\sinh^3(u)}. \nonumber
\end{eqnarray}
Combining these we obtain the following contribution from \eqref{euler2}
\begin{equation}
- \frac{\Omega_3}{4 \pi^2} \left (\cosh^3(\bar{u}) - 3 \cosh  (\bar{u}) \right ) = - \frac{\Omega_3}{32 \pi^2} \left ( e^{3 \bar{u}} - 9 e^{\bar{u}}  + \cdots \right )
\end{equation}
where in the second 	expression we have dropped all terms that go to zero as $\bar{u} \rightarrow \infty$. 

Combining bulk and boundary terms we obtain
\begin{equation}
\chi ({\cal S}) = 1,
\end{equation}
which is indeed the Euler invariant for a half ball. 

\bigskip

Let us now turn to the computation of the renormalized entanglement entropy. The regulated bulk contribution is proportional to the regulated volume of the entangling surface
\begin{equation}
\frac{\Omega_3}{4 G_6} \left (  \frac{2}{3} + \frac{1}{24} e^{3 \bar{u}} - \frac{3}{8} e^{\bar{u}} + \cdots \right ) 
\end{equation}
where the ellipses denote terms that vanish as $\bar{u} \rightarrow \infty$. The first counterterm gives
\begin{equation}
- \frac{\Omega_3}{4 G_6} \left ( \frac{1}{24} e^{3 \bar{u}} - \frac{1}{8} e^{\bar{u}} + \cdots \right ) 
\end{equation}
while the second counterterm gives 
\begin{equation}
\frac{\Omega_3}{4 G_6} \left ( \frac{1}{4} e^{\bar{u}} + \cdots \right ).
\end{equation}
The counterterms, as expected, remove divergent contributions while not adding further finite contributions and thus 
\begin{equation}
S( {\cal S}) = \frac{\pi^2}{3 G_{6}} \equiv  \frac{\pi^2}{3 G_{6}}  \chi (\Sigma),
\end{equation}
i.e. the renormalized entanglement entropy is proportional to the Euler invariant, as shown in \eqref{structure}, with the coefficient of proportionality being the F quantity in the dual CFT$_5$. 

\bigskip

Next let us consider the variation of the entanglement entropy under linear perturbations around the spherical entangling surface in AdS. Since the Weyl curvature and the extrinsic curvatures are zero at leading order, only terms linear in the curvatures can contribute to the first variation of the entropy. From \eqref{eq:A6} we obtain
\be
\delta S = -\frac{1}{12 G_6} \delta{\cal W}
\ee
where ${\cal W}$ is linear in the Weyl curvature and is defined in \eqref{eq:wsigma}. 

As we will show in a future work \cite{FirstLaw}, this expression allows for an elegant derivation of the first law of entanglement entropy, generalizing the discussions in \cite{Faulkner:2013ica}. Since we work with renormalized quantities, we do not need to assume specific fall off conditions for metric perturbations; the perturbations can be non-normalizable as well as normalizable. It is straightforward to relate $\delta {\cal W}$ 
to the renormalized stress tensor defined in \cite{deHaro2001} and hence to the variation in the energy.

	\section{Conclusions and outlook} \label{sec:five}
	
In this paper we have shown that the renormalized area of static minimal surfaces in asymptotically locally AdS$_{2n}$ spacetimes can be expressed in terms of the Euler invariant and renormalized curvature invariants. It is perhaps unsurprising that renormalized integrals of extrinsic and intrinsic curvature invariants arise. Indeed, renormalized curvature integrals on asymptotically locally hyperbolic manifolds have been considered in the mathematics literature; see for example \cite{Albin}. 

There is however a key difference between our definitions of renormalized curvature invariants and those in the mathematics literature. Here we follow the standard holographic renormalization approach, identifying explicit boundary counterterms. By contrast, the mathematics literature identifies the ``renormalized'' integrals as the finite terms in a regulated expansion around the conformal boundary. While the latter gives equivalent results when counterterms do not make finite contributions, there will generically be finite contributions from counterterms (for example, if we add matter or gauge fields in the bulk). 

\bigskip

Our results can be used to infer certain bounds on the renormalized entanglement entropy for given topology. Earlier discussions on bounds on renormalized entanglement entropy in asymptotically locally AdS$_4$ spacetimes using inverse mean curvature flow techniques can be found in \cite{Fischetti:2016fbh}. For entangling surfaces of disk topology in AdS$_4$, the bound is given in \eqref{bound1}: the entanglement entropy is negative and the absolute value of the renormalized entanglement entropy is minimised for the disk, which has zero extrinsic curvature. 

Note that the expression \eqref{bound1} is closely analogous to the Willmore energy ${\cal E}_w$, which measures how much a closed two surface $\Sigma$ embedded into {R}$^3$ deviates from the round two sphere: 
\be
{\cal E}_w  = \int_{\Sigma} d^2 x |K|^2  - 2 \pi \chi (\Sigma)
\ee
The Willmore energy is positive semi-definite and zero for a round two sphere. 

For entangling surfaces that are topologically disks in asymptotically locally AdS$_4$ manifolds (cf pure AdS$_4$), there is no such bound: from \eqref{simp1}, the Weyl curvature term is not negative definite. Indeed, if one considers linearized perturbations around AdS$_4$, one can show that this term is positive for all metric perturbations that give rise to positive energy \cite{FirstLaw}. 

\bigskip

Now let us turn to the renormalized entanglement entropy for asymptotically locally AdS$_6$ spacetimes. From \eqref{eq:A6}, this reduces in AdS$_6$ to 
\begin{eqnarray}
S (\Sigma) &=& \frac{ \pi^2}{3 G_6} \chi (\Sigma) - \frac{1}{24 G_6} {\cal H}(\Sigma) \label{bounda6} \\ 
&& - \frac{1}{96 G_6} \int_{\Sigma} d^4 x \sqrt{g} \left ( H^2 - 4 H_{\mu \nu} H^{\mu \nu} + H_{\mu \nu \rho \sigma} H^{\mu \nu \rho \sigma} \right ). \nonumber
\end{eqnarray}
Even restricting to entangling surfaces of fixed topology, this expression does not seem to have bounds, in accordance with the discussions of higher dimensional Willmore functionals in \cite{Zhang:2017lcd,Graham:2017bew}. 

For example, consider perturbations around a spherical entangling surface in AdS$_6$; the change in the renormalized entanglement entropy is 
\be
\delta S = -\frac{1}{24 G_6} \delta {\cal H} = -\frac{1}{24 G_6} \int_{\Sigma} d^4 x \sqrt{g} \delta H + \frac{1}{24 G_6} 
\int_{\partial \Sigma} d^3 x \sqrt{h} \left ( \delta k_{ij} \delta k^{ij} - \frac{1}{3} (\delta k)^2 \right ),
\ee
i.e. it is quadratic in the extrinsic curvature. The bulk curvature integrand is non-positive but the renormalized curvature invariant does not manifestly have any negativity bounds. Hence, in 3d holographic CFTs, disk regions minimise the magnitude of the renormalized entanglement entropy in the conformal vacuum while in 5d holographic CFTs spherical regions do not necessarily do so. It would be interesting to understand the implications of this directly from field theory. 

\bigskip

Throughout this paper we have been considering static RT entangling surfaces \cite{Ryu:2006bv} although our analysis of the Chern-Gauss-Bonnet integrals is applicable to generic asymptotically locally AdS manifolds. It would be interesting to extend our analysis of renormalized entanglement entropy to HRT surfaces \cite{Hubeny:2007xt}. 

While we have focussed on connected entangling regions, our expressions for renormalized entanglement entropy are equally applicable to disconnected regions.  If we consider $n$ widely separated disk/spherical entangling regions in pure AdS then from \eqref{bound1} and \eqref{bounda6} the entanglement entropy is proportional to $n/G$. It would be interesting to explore how the renormalized entanglement entropy changes as these regions approach each other and intersect. In AdS$_4$ the extremum once all regions intersect would be a single disk region with entropy proportional to $1/G$ and the renormalized entanglement entropy may satisfy monotonicity properties under deformations of disconnected regions into a single connected region.

	\section*{Acknowledgements}
	
	This work is funded by the STFC grant ST/P000711/1. This project has received funding and support from the European Union's Horizon 2020 research and innovation programme under the Marie Sklodowska-Curie grant agreement No 690575. MMT would like to thank the Kavli Institute for the Physics and Mathematics of the Universe and the Indian Institute of Technology in Kanpur for hospitality during the completion of this work. 	
	
			\appendix
			
	\section{Appendix: Notation and terminology}
	
	In this appendix we collect together notation and terminology. 
	
	We denote the curvature of the asymptotically locally AdS manifold ${\cal M}$ (metric $G_{\mu \nu}$) with boundary $\partial {\cal M}$ (metric $\gamma_{\alpha\beta}$) as $R_{\mu \nu \rho \sigma}$. The intrinsic curvature of the boundary of $\partial {\cal M}$ is denoted by $\hat{R}_{\alpha\beta\gamma\delta}$. The entangling surface ${\Sigma}$ with metric $g$ has boundary ${\partial \Sigma}$ with metric $h$. The intrinsic curvature of the surface $\Sigma$ is denoted ${\cal R}_{ijkl}$. The intrinsic curvature of the surface $\partial\Sigma$ is denoted $\bar{R}_{ijkl}$. 

We also need to distinguish between four distinct extrinsic curvatures: the extrinsic curvatures of $\Sigma$ embedded into $M$ ($K^s$), of $\partial \Sigma$ embedded into $\Sigma$ (${\cal K}$), of $\partial \Sigma$ embedded into $\partial M$ ($k^s$), of $\partial\Sigma$ embedded into $M$ ($k^{\perp}$) orthogonal to $\partial M$ and of $\partial M$ embedded into $M$ ($\mathscr{K}$), where $s=1,2$ denote the normals to the entangling surface with the boundary condition $K^s=k^s$ on $\partial\Sigma$. Note that we write $k^{(2)}=k$ and in the static case $K^{(1)}=k^{(1)}=0$.
			
			\section{Appendix: Asymptotic analysis}
	The boundary metric $\gamma$ has a Fefferman-Graham expansion therefore the metrics $g$ and $h$ on $\Sigma$ and $\partial\Sigma$ have their own Fefferman-Graham expansion,
\begin{align}
g_{ij}=\frac{1}{z^2}\bar{g}_{ij}=\frac{1}{z^2}\left(\bar{g}^{(0)}_{ij}+z^2\bar{g}^{(2)}_{ij}+\cdots\right)
\end{align}
and
\begin{align}
h_{ij}=\frac{1}{z^2}\bar{h}_{ij}=\frac{1}{z^2}\left(\bar{h}^{(0)}_{ij}+z^2\bar{h}^{(2)}_{ij}+\cdots\right).
\end{align}
Hence $\bar{\alpha}$ has an even power series of $z$, 
\begin{align}
\bar{\alpha}=\bar{\alpha}^{(0)}+z^2\bar{\alpha}^{(2)}+O(z^4).
\end{align}
Then 
\begin{align}
\bar{\beta}=\frac{2z^2\bar{\alpha}^{(2)}}{\bar{\alpha}^{(0)}}+O(z^4).
\end{align}
Similarly, for $e^a_i$ has an even power series of $z$,
\begin{align}
e^a_{\;i}=\frac{1}{z}\bar{e}^a_{\;i}=\frac{1}{z}\left(\bar{e}^{(0)a}_{\;\;\;\;\;\;i}+z^2\bar{e}^{(2)a}_{\;\;\;\;\;\;i}+\cdots\right)
\end{align}

The extrinsic curvature $k^{\perp}$ of $\partial\Sigma$ pointing out of $\partial M$,
\begin{align}
k^{\perp}_{ab}&=(\nabla m)_{ab}\nonumber\\
&=-\Gamma^{m}_{ab}\nonumber\\
&=e_{(b}^{\;\;k}m\cdot\partial e_{a)k}\nonumber\\
&=z\partial_z\left(\frac{1}{z}\right)e_{(b}^{\;\;k}\bar{e}_{a)k}+e_{(b}^{\;\;k}\partial_z\bar{e}_{a)k}\nonumber\\
&=-\delta_{ab}+2\bar{e}_{\;\;\;(b}^{(0)\;\;k}\bar{e}^{(2)}_{\;\;\;a)k}+O(z^4)\nonumber\\
k^{\perp}_{ab}&=-\delta_{ab}+O(z^4)
\end{align}
where we used the fact $\bar{e}_{(b}^{\;k}\bar{e}_{a)k}=\delta_{ab}$ which implies $\bar{e}_{\;\;\;b}^{(0)\;k}\bar{e}_{\;\;\;ak}^{(2)}+\bar{e}_{\;\;\;b}^{(2)\;k}\bar{e}^{(0)}_{\;\;\;ak}=0$. Transforming back to coordinate basis,
\begin{align}
k^{\perp}_{ij}&=-h_{ij}+O(z^2)
\end{align}
The other extrinsic curvature $k$ of $\partial\Sigma$ lying within $\partial M$,
\begin{align}
k_{ab}&=e_{(b}^{\;\;\;k}\bar{n}\cdot\partial e_{a)k}\sim O(z).
\end{align}
In coordinate basis,
\begin{align}
k_{ij}\sim O(z^{-1})
\end{align}
\subsection{Asymptotic analysis for bulk Euler density}
Starting from the definition of the extrinsic curvature 
\begin{equation}
K^{(2)}_{\mu\nu}=(h^{\rho}_{\mu}+n^3_{\mu}n^{3\;\rho})(h^{\sigma}_{\nu}+n^3_{\nu}n^{3\;\sigma})\nabla_{\rho}(A\bar{n}_\sigma+A^{\perp}m_{\sigma}).
\end{equation}
Expanding the bracket and grouping the terms tangent and normal to $\partial\Sigma$
\begin{align}
K^{(2)}_{\mu\nu}=&\;Ak_{\mu\nu}+A^{\perp}k^{\perp}_{\mu\nu}+h^{\rho}_{\mu}n^3_{\nu}\Big(-A^{\perp}\partial_{\rho}A+A\partial_{\rho}A^{\perp}+m^{\sigma}\nabla_{\rho}\bar{n}_{\sigma}\\
&-AA^{\perp}(\bar{n}^{\sigma}\nabla_{\sigma}\bar{n}_{\rho}-m^{\sigma}\nabla_{\sigma}m_{\rho})-{A^{\perp}}^2[\bar{n},m]_{\rho}\Big)+n^3_{\mu}n^3_{\nu}n^3_{\sigma}[n^3,n^2]^{\sigma}.\nonumber
\end{align}
As mentioned before, $\partial_iA=0$ and $[\bar{n},m]\in N\partial\Sigma$ so the terms with one index tangent and one index normal to $\partial\Sigma$ vanish. Then expand the Lie bracket of $n^2,n^3$ and use the relation $\partial_{\rho}A^{\perp}=-\frac{A}{A^{\perp}}\partial_{\rho}A$, the expression simplifies to
\begin{align}
K^{(2)}_{\mu\nu}=&\;Ak^{(2)}_{\mu\nu}+A^{\perp}k^{\perp}_{\mu\nu}+n^3_{\mu}n^3_{\nu}\Big[\partial_{\rho}A\Big(\bar{n}^{\rho}+m^{\rho}(-\frac{A}{A^{\perp}}+A^2-AA^{\perp})\Big)-\Big(1-\bar{\beta}\Big)A^{\perp}\Big].
\end{align}
Finally defining the coefficient in the $n^3_{\mu}n^3_{\nu}$ component in $K^{(2)}_{\mu\nu}$
\begin{align}
L\vcentcolon = \partial_{\rho}A\Big(\bar{n}^{\rho}+m^{\rho}(-\frac{A}{A^{\perp}}+A^2-AA^{\perp})\Big)-\Big(1-\bar{\beta}\Big)A^{\perp},
\end{align} the extrinsic curvature has tangent components,
\begin{align}
K^{(2)}_{ij}=Ak_{ij}+A^{\perp}k^{\perp}_{ij}
\end{align}
and normal components,
\begin{align}
K^{(2)}_{\bar{n}m}=-AA^{\perp}L,\;\;\;\;K^{(2)}_{\bar{n}\bar{n}}=-{A^{\perp}}^2L,\;\;\;\;K^{(2)}_{mm}=-A^2L.
\end{align}
Using $(\ref{eq:AA})$, expanding L in $z$,
\begin{align}
L&=-\frac{z^3A^{\perp}_{(0)}\partial_{\bar{f}}A^{\perp}_{(0)}}{\bar{\alpha_0}}-z^2{A^{\perp}_{(0)}}^2\left(-\frac{1}{zA^{\perp}_{(0)}}+z\left(1+\frac{A^{\perp}_{(2)}}{{A^{\perp}_{(0)}}^2}-\frac{{A^{\perp}_{(0)}}^2}{2}\right)\right)\\
&-z{A^{\perp}_{(0)}}-z^3{A^{\perp}_{(2)}}+\frac{2z^3\bar{\alpha}_2A^{\perp}_{(0)}}{\bar{\alpha}_0}+O(z^5)\nonumber\\
L&=-z^3\left(\frac{A^{\perp}_{(0)}\partial_{\bar{f}}A^{\perp}_{(0)}}{\bar{\alpha_0}}+{A^{\perp}_{(0)}}^2+2A^{\perp}_{(2)}-\frac{{A^{\perp}_{(0)}}^4}{2}-\frac{2\bar{\alpha}_2A^{\perp}_{(0)}}{\bar{\alpha}_0}\right)+O(z^5).\nonumber
\end{align}
Therefore trace of the extrinsic curvature,
\begin{align}
K^{(2)}&=Ak+A^{\perp}k^{\perp}+L\sim O(z)\label{eq:TK2}
\end{align}
and trace of the product of extrinsic curvature,
\begin{align}
K^{(2)}_{\mu\nu}K^{(2)\mu\nu}&=A^2k_{\mu\nu}k^{\mu\nu}+{A^{\perp}}^2k^{\perp}_{\mu\nu}k^{\perp\mu\nu}+2AA^{\perp}k_{\mu\nu}k^{\perp\mu\nu}+L^2\label{eq:TK2K2}\\
K^{(2)}_{\mu\nu}K^{(2)\mu\nu}&=k_{\mu\nu}k^{\mu\nu}+3{A^{\perp}}^2-2A^{\perp}k+O(z^4)\sim O(z^2).\nonumber
\end{align}
From $(\ref{eq:TK2})$ and $(\ref{eq:TK2K2})$ we observed that up to $O(z^4)$ only $H$ contains divergent integrals. Further expanding in $z$,
\begin{align}
{K^{(2)}}^2&=k^2+9z^2{A^{\perp}_{(0)}}^2-6z{A^{\perp}_{(0)}}k+O(z^4)\\
{K^{(2)}}^2&=z^2\left(\bar{k}_0^2+9{A^{\perp}_{(0)}}^2-6{A^{\perp}_{(0)}}\bar{k}_0\right)+O(z^4)\nonumber
\end{align}
and
\begin{align}
K^{(2)}_{ij}K^{(2)ij}&=k_{ij}k^{ij}+3z^2{A^{\perp}_{(0)}}^2-2zA^{\perp}_{(0)}k+O(z^6)\\
K^{(2)}_{ij}K^{(2)ij}&=z^2\left(\bar{k}_{0ij}\bar{k}_0^{ij}+3{A^{\perp}_{(0)}}^2-2A^{\perp}_{(0)}\bar{k}_0\right)+O(z^6).\nonumber
\end{align}
The leading order term in the Taylor expansion of the extrinsic curvature $\bar{k}_{(0)}$ can be written in terms of the leading order term in the Fefferman-Graham expansion of boundary induced metric  $\bar{h}^{(0)}_{ij}$
\begin{align}
\bar{k}_{(0)}=\left[z^{-1}k\right]_{z=0}=\frac{1}{\bar{\alpha}^{(0)2}}\bar{h}^{(0)ij}\partial_{\bar{f}}\bar{h}^{(0)}_{ij}
\end{align} 
and 
\begin{align}
\bar{k}_{(0)ij}\bar{k}_{(0)}^{ij}=\left[z^{-2}k_{ij}k^{ij}\right]_{z=0}=\frac{1}{\bar{\alpha}^{(0)2}}\bar{h}^{(0)ik}\partial_{\bar{f}}\bar{h}^{(0)}_{ij}\bar{h}^{(0)jl}\partial_{\bar{f}}\bar{h}^{(0)}_{kl}.
\end{align}

\subsection{Asymptotic analysis for boundary Euler density}
Now	consider the regulated divergences in the boundary terms in the Euler characteristic. 
From power counting the dominant order of each term in $(\ref{eq:FdE4})$, only the following terms contain divergent integral
\begin{align}
\partial E_4&=A k^{\perp}\left(W_{1\bar{n}1\bar{n}}+W_{1m1m}-W_{\bar{n}m\bar{n}m}\right)\\
&-A k^{\perp\;ij}\left(-W_{i1j1}+W_{imjm}+W_{i\bar{n}j\bar{n}} \right)\nonumber\\
&+Ak^{\perp}-\left(\frac{A}{2}-\frac{A^3}{6}\right){k^{\perp}}^3-\left(A-\frac{A^3}{3}\right) k^{\perp}_{ij}k^{\perp\;jk}k^{\perp\;i}_{\;k}\nonumber\\
&-\left(-\frac{3A}{2}+\frac{A^3}{2}\right)k^{\perp}k^{\perp}_{ij}k^{\perp\;ij}\nonumber\\
&-A^{\perp}k-\left(-\frac{A^{\perp}}{2}+\frac{A^2A^{\perp}}{2}\right)k {k^{\perp}}^2-\left(\frac{A^{\perp}}{2}-\frac{A^2A^{\perp}}{2}\right)k k^{\perp}_{ij}k^{\perp\;ij}\nonumber\\
&-\left(-A^{\perp}+A^2A^{\perp}\right)k_{ij}k^{\perp\;jk}k^{\perp \;i}_{\;k}-\left(A^{\perp}-A^2A^{\perp}\right)k^{\perp}k_{ij}k^{\perp\;ij}\nonumber\\
&-\frac{A}{2}k^2k^{\perp}+\frac{A}{2}k_{ij}k^{ij}k^{\perp}+Ak_{ij}k^{jk}k^{\perp \;i}_{\;k}+A k k_{ij}k^{\perp\;ij}.\nonumber
\end{align}
Simplifying by substituting the leading order term of $k^{\perp}_{ij}=-h_{ij}$,
\begin{align}
\partial E_4^{div}&=-3\left(W_{1\bar{n}1\bar{n}}+W_{1m1m}-W_{\bar{n}m\bar{n}m}\right)\\
&+h^{ij}\left(-W_{i1j1}+W_{imjm}+W_{i\bar{n}j\bar{n}} \right)\nonumber\\
&-3A+\frac{27A}{2}-\frac{9A^3}{2}+3A-A^3+\frac{27A}{2}+\frac{9A^3}{2}\nonumber\\
&-A^{\perp}k+\frac{9A^{\perp}}{2}k-\frac{9A^2A^{\perp}}{2}k-\frac{3A^{\perp}}{2}k+\frac{3A^2A^{\perp}}{2}k\nonumber\\
&+A^{\perp}k-A^2A^{\perp}k-3A^{\perp}k+3A^2A^{\perp}k\nonumber\\
&+\frac{3}{2}k^2-\frac{3}{2}k_{ij}k^{ij}+k_{ij}k^{ij}-Ak^2.\nonumber
\end{align}
Expanding the induced metric and using tracelessness of the Weyl tensor,
\begin{align}
\partial E_4^{div}&=-\left(W_{1\bar{n}1\bar{n}}+W_{1m1m}-W_{\bar{n}m\bar{n}m}+A^3+A^{\perp}k-\frac{1}{2}k^2+\frac{1}{2}k_{ij}k^{ij}\right).
\end{align}

To look at the detail asymptotic behaviour of the Weyl tensor we need the expression of Riemann tensor in Fefferman-Graham gauge. Particularly for $W_{1\bar{n}1\bar{n}},W_{1m1m},W_{\bar{n}m\bar{n}m}$ 
\begin{align}
R_{tztz}&=-\frac{1}{z^4}\bar{\gamma}_{tt}+\frac{1}{4z^2}\left(-2\bar{\gamma}_{tt}''+\bar{\gamma}_{t\mu}'\bar{\gamma}^{\mu\nu}\bar{\gamma}_{\nu t}'\right)+\frac{1}{2z^3}\bar{\gamma}_{tt}'\label{eq:Rtztz}\\
R_{\bar{f}z\bar{f}z}&=-\frac{1}{z^4}\bar{\gamma}_{\bar{f}\bar{f}}+\frac{1}{4z^2}\left(-2\bar{\gamma}_{\bar{f}\bar{f}}''+\bar{\gamma}_{\bar{f}\mu}'\bar{\gamma}^{\mu\nu}\bar{\gamma}_{\nu \bar{f}}'\right)+\frac{1}{2z^3}\bar{\gamma}_{\bar{f}\bar{f}}'\label{eq:Rfzfz}\\
R_{t\bar{f}t\bar{f}}&=\frac{1}{z^4}\left(\bar{\gamma}_{t\bar{f}}^2-\bar{\gamma}_{tt}\bar{\gamma}_{\bar{f}\bar{f}}\right)+\hat{R}_{t\bar{f}t\bar{f}}[\gamma]+\frac{1}{4z^2}\left({\bar{\gamma}_{t\bar{f}}}^{\prime 2}-\bar{\gamma}_{tt}'\bar{\gamma}_{\bar{f}\bar{f}}'\right)\\
&+\frac{1}{2z^3}\left(\bar{\gamma}_{tt}'\bar{\gamma}_{\bar{f}\bar{f}}+\bar{\gamma}_{tt}\bar{\gamma}_{\bar{f}\bar{f}}'-2\bar{\gamma}_{t\bar{f}}\bar{\gamma}_{t\bar{f}}'\right)\nonumber\\
R_{t\bar{f}tz}&=\frac{1}{2z^2}\left(D_t\bar{\gamma}_{t\bar{f}}'-D_{\bar{f}}\bar{\gamma}_{tt}'\right)
\end{align}
where ${}'=\partial_z$ and $D$ is the covariant derivative on $\partial M$. Note in Fefferman-Graham gauge, the derivatives of the metric scale as $\bar{\gamma}_{\mu\nu}'\sim O(z)$ and $\bar{\gamma}_{\mu\nu}''\sim O(1)$. In our gauge 
\begin{align}
W_{1212}&=z^4{A^{\perp}}^2R_{tztz}+\frac{z^4A^2}{\bar{\alpha}^2}R_{t\bar{f}t\bar{f}}+\frac{2AA^{\perp}}{\bar{\alpha}}R_{t\bar{f}tz}\\
W_{1212}&=z^2\left(\frac{z^2}{\bar{\alpha}_0^2}\hat{R}_{t\bar{f}t\bar{f}}-\frac{2\bar{\alpha}_2}{\bar{\alpha}_0}\right)+O(z^4)\nonumber
\end{align}
and
\begin{align}
W_{1\bar{n}1\bar{n}}&=\frac{z^4}{\bar{\alpha}^2}R_{t\bar{f}t\bar{f}}-1\\
W_{1\bar{n}1\bar{n}}&=z^2\left(\frac{z^2}{\bar{\alpha}_0^2}\hat{R}_{t\bar{f}t\bar{f}}-\frac{2\bar{\alpha}_2}{\bar{\alpha}_0}\right)+O(z^4)\nonumber\\
W_{1\bar{n}1\bar{n}}&\sim W_{1212}\nonumber
\end{align}
where the equivalence is up to order $O(z^4)$. Similarly,
\begin{align}
W_{\bar{n}m\bar{n}m}&=\frac{z^4}{\bar{\alpha}^2}R_{\bar{f}z\bar{f}z}+1\label{Wnm0}\\
W_{\bar{n}m\bar{n}m}&=1+\frac{z^4}{\bar{\alpha_0^2}}\left(-\frac{\bar{\alpha}_0^2}{z^4}-\frac{2\bar{\alpha}_0\bar{\alpha}_2}{2z^2}+\frac{2\bar{\alpha}_0\bar{\alpha}_2}{z^3}\right)+O(z^4)\nonumber\\
W_{\bar{n}m\bar{n}m}&\sim 0\nonumber
\end{align}
and 
\begin{align}
W_{1m1m}&=z^4R_{tztz}-1\\
W_{1m1m}&\sim 0.\nonumber
\end{align}
Although for our gauge $\bar{\gamma}_{tt}'=0$, in non-static boundary one can replace the normalization constant of the spacelike unit normal, $\bar{\alpha}$, to the normalization constant of timelike orthonormal basis, $\alpha_{\tau}$, and $W_{1m1m}$ should also vanish up to $O(z^4)$.

The minimal condition for $\Sigma$ is equivalent to having a vanishing trace for the extrinsic curvature $K^{(2)}=0$; the vanishing of Lie derivative of the volume form of $\Sigma$ with respect to the normal $n^2$, $(\ref{eq:TK2})$ implies

\begin{align}
A^{\perp}&=\frac{-Ak-L}{k^{\perp}}\\
A^{\perp}_{(0)}&=\frac{\bar{k}_{(0)}}{3}\nonumber.
\end{align}

\section{Chern Gauss Bonnet Formula}
The construction of the Chern Gauss Bonnet Formula follows from \cite{Chern:10.2307/1969302,Dowker:1989ue,li2011gaussbonnetchern}. The connection 1-form $\omega_{ab}$ is defined by
\begin{align}
\omega^a_b=\Gamma^a_{bc}e^c,\quad de^a=-\omega^a_{\;b}\land e^b,\quad de_b=\omega^a_{\;b}e_a
\end{align}
and the curvature 2-form $\Omega_{ab}$ is defined by
\begin{align}
\Omega_{\;b}^a= \frac{1}{2}\mathcal{R}^a_{\;bcd}e^c\land e^d,\quad \Omega^a_b=d\omega^a_{\;b}+\omega^a_{\;c}\land\omega^c_{\;b}.
\end{align}
Consider a vector field $X$ of a d dimensional manifold $M$ with zeros of the vector field $I\subset M$. For a $d-1$ sphere bundle $\pi:SM\rightarrow M$, one can identify a map, by the normalized vector field, from the $M\setminus I$ to $SM$ such that $\hat{X}\in \Gamma\left(M\setminus I,SM\right)$. The d form $\Omega_p \in \land^dT^*_pM$,
\begin{align}
\Omega_p=\frac{1}{2^d\pi^{\frac{d}{2}}\left(\frac{d}{2}\right)!}\epsilon_{a_1\cdots a_{d}}\Omega^{a_1a_2}\land\cdots\land\Omega^{a_{d-1}a_d},
\end{align}
is exact when pullback to $SM$
\begin{align}
\pi^*\Omega=-d\Pi.
\end{align} 
The exact form on $SM$ is
\begin{align}
\Pi=\frac{1}{\pi^{\frac{d}{2}}}\sum_{k=0}^{\frac{d}{2}-1}\frac{1}{1\cdot3\cdots(d-2k-1)k!2^{\frac{d}{2}+k}}\Phi_k
\end{align}
The $\Phi_k$ are $d-1$ forms on $SM$
\begin{align}
\Phi_k=\epsilon_{a_1\cdots a_{d}}u^{a_1}\theta^{a_2}\land\cdots\land\theta^{a_{d-2k}}\land\pi^*\Omega^{a_{d-2k+1}a_{d-2k+2}}\land\cdots\land\pi^*\Omega^{a_{d-1}a_d}
\end{align}
where $u$ is an unit tangent vector of $M$ pullback to $SM$ and the 1-form $\theta$ is defined by 
\begin{align}
\theta^a=du^a+u^b\pi^*\omega_{\;b}^{a}.
\end{align} 
Then using Stoke's theorem and the fact $\hat{X}^*\pi^*=id_M$
\begin{align}
\int_M \Omega &= \int_{\hat{X}\left(M\right)}\pi^*\Omega=-\int_{\hat{X}\left(M\right)}d\Pi\\
&=-\int_{\hat{X}\left(M\setminus \cup_{x\in I}B_x\right)}d\Pi-\int_{\hat{X}\left(\cup_{x\in I}B_x\right)}d\Pi\nonumber\\
&=-\int_{\hat{X}\left(\partial M\right)}\Pi+\int_{\hat{X}\left(\cup_{x\in I}\partial \bar{B}_x\right)}\Pi-\int_{\cup_{x\in I}B_x}\Omega\nonumber\\
&\stackrel{\lim\limits_{r \to 0}}{=}-\int_{\hat{X}\left(\partial M\right)}\Pi+\int_{\hat{X}\left(\cup_{x\in I}\partial \bar{B}_x\right)}\Pi\nonumber\\
&=-\int_{\partial M}\hat{X}^*\Pi+\int_{\cup_{x\in I}\partial \bar{B}_x}\hat{X}^*\Pi\nonumber\\
&=-\int_{\partial M}\hat{X}^*\Pi+\sum_{x\in I}deg_x(\hat{X})\int_{\partial \bar{B}_x}\Pi\nonumber\\
&=-\int_{\partial M}\hat{X}^*\Pi+\sum_{x\in I}index_x(\hat{X})\nonumber,
\end{align}
rearranging the above equation and apply the Poincare-Hopf theorem we get the Chern-Gauss-Bonnet formula
\begin{align}
\int_M \Omega +\int_{\partial M}\hat{X}^*\Pi=\chi(M).
\end{align}

\subsection{d=4 Riemannian manifold with boundary}
Our entangling surface is a 4 dimensional Riemannian manifold.
\begin{align}
\Omega&=\frac{1}{32\pi^2}\epsilon_{abcd}\Omega^{ab}\land\Omega^{cd}\\
\Omega&=\frac{1}{128\pi^2}\epsilon_{abcd}\mathcal{R}^{ab}_{\;\;\;ef}\mathcal{R}^{cd}_{\;\;\;gh}e^ee^fe^ge^h\nonumber\\
\Omega&=\frac{1}{128\pi^2}\epsilon_{abcd}\epsilon^{efgh}\mathcal{R}^{ab}_{\;\;\;ef}\mathcal{R}^{cd}_{\;\;\;gh}e^1e^2e^3e^4\nonumber\\
\Omega&=\frac{1}{32\pi^2}\left(\mathcal{R}^{ab}_{\;\;\;cd}\mathcal{R}^{cd}_{\;\;\;ab}-4\mathcal{R}^a_{\;b}\mathcal{R}^b_{\;a}+\mathcal{R}^2\right)dV
\end{align}
We can choose the vector field $\hat{X}$ to be the inward pointing unit normal of the boundary $n=e_4$ then 
\begin{align}
n^*\Pi=\frac{1}{\pi^2}\left(\frac{1}{12}n^*\Phi_0+\frac{1}{8}n^*\Phi_1\right).
\end{align}
The 3-form $\Phi_0$ is explicitly written in terms of extrinsic curvature of the boundary $\partial M$ as,
\begin{align}
n^*\Phi_0&=\epsilon_{4ijk}\omega^{i}_{\;4}\land\omega^{j}_{\;4}\land\omega^{k}_{\;4}\\
n^*\Phi_0&=\epsilon_{ijk}\epsilon^{lpq}\mathcal{K}^i_l\mathcal{K}^j_p\mathcal{K}^k_qe^1e^2e^3\nonumber\\
n^*\Phi_0&=\left(\mathcal{K}^3-3\mathcal{K}\textrm{Tr}(\mathcal{K}^2)+2\textrm{Tr}(\mathcal{K}^3)\right)dS\label{eq:phi0}
\end{align}
to get to the second line we used $\omega^{i}_{\;4}$ on the boundary is related to the extrinsic curvature
\begin{align}
\omega^{i}_{\;4}=-\mathcal{K}^i_je^j.
\end{align}
Similarly the 3-form $\Phi_1$ is written in terms of the intrinsic curvature of $M$ and the extrinsic curvature of $\partial M$,
\begin{align}
n^*\Phi_1&=\epsilon_{4ijk}\omega^i_{\;4}\land\Omega^{jk}\\
n^*\Phi_1&=\frac{1}{2}\epsilon_{ijk}\epsilon^{lpq}\mathcal{K}^i_l\mathcal{R}^{jk}_{\;\;\;pq}e^1e^2e^3\nonumber\\
n^*\Phi_1&=\mathcal{K}\mathcal{R}-2\mathcal{K}\mathcal{R}_{ab}n^an^b-2\mathcal{K}^{ab}\mathcal{R}_{ab}+2\mathcal{R}_{abcd}\mathcal{K}^{ac}n^bn^d\label{eq:phi1}
\end{align}
Combining the $(\ref{eq:phi0})$ and $(\ref{eq:phi1})$ and using the coordinate on $\Sigma$, $x^i,\;i=1,2,3,4$, we recover $(\ref{euler})$
\begin{align}
\int_{\Sigma}\Omega=\frac{1}{32\pi^2}\int_{\Sigma}d^4x\left ( {\cal R}^{ijkl} {\cal R}_{ijkl} - 4 {\cal R}_{ij} {\cal R}^{ij} + {\cal R}^2 \right)
\end{align}
and $(\ref{euler2})$
\begin{align}
\int_{\partial \Sigma} n^* \Pi=\frac{1}{4 \pi^2} \int_{\partial \Sigma} d^3x \sqrt{h} \bigg( & {\cal R}_{ijkl} {\cal K}^{ik} n^{j} n^k - {\cal R}^{ij} {\cal K}_{ij} - {\cal K} {\cal R}_{ij} n^i n^j + \frac{1}{2} {\cal K} {\cal R} \\
 &+ \frac{1}{3} {\cal K}^3  - {\cal K} {\rm Tr} ( {\cal K} ^2) +  \frac{2}{3} {\rm Tr} ( {\cal K}^3) \bigg) \nonumber.
\end{align}
Note that $(\ref{euler2})$ is independent of the orientation of the normal $n$ because the extrinsic curvature is only defined by the outward pointing normal. 
	\bibliography{refs}
	
	
	

\end{document}